\documentclass[aps,pre,reprint,twocolumn,showpacs,preprintnumbers,groupedaddress,floatfix,amsmath,amssymb]{revtex4-1}

\usepackage{graphicx}
\usepackage{color}

\newcommand{\physrep}{Phys.~Rep.} 
\newcommand{\apjl}{Astrophys.~J.~Lett.} 
\newcommand{\aap}{Astron.~\&~Astrophys.} 
\newcommand{\araa}{Annu.~Rev.~Astron.~Astrophys.} 
\newcommand{\mnras}{Mon.~Not.~R.~Astron.~Soc.} 

\newcommand{\Pm}{\text{Pm}}
\newcommand{\Rey}{\text{Re}}
\newcommand{\Rm}{\text{Rm}}
\newcommand{\lnu}{\ell_{\nu}}
\newcommand{\leta}{\ell_{\eta}}
\newcommand{\tabhe}{\parbox[0pt][2em][c]{0cm}{}}

\begin{document}

\title{The Small-Scale Dynamo at Low Magnetic Prandtl Numbers}

\author{Jennifer Schober}
 \email{schober@stud.uni-heidelberg.de}
 \affiliation{Universit\"at Heidelberg, Zentrum f\"ur Astronomie, Institut f\"ur Theoretische Astrophysik, Albert-\"Uberle-Strasse\ 2, D-69120 Heidelberg, Germany}
\author{Dominik Schleicher}
 \email{dschleic@astro.physik.uni-goettingen.de}
 \affiliation{Institut f\"ur Astrophysik, Georg-August-Universit\"at G\"ottingen, Institut f\"ur Astrophysik, Friedrich-Hund-Platz 1, D-37077 G\"ottingen, Germany}
\author{Stefano Bovino}
 \email{sbovino@astro.physik.uni-goettingen.de}
 \affiliation{Institut f\"ur Astrophysik, Georg-August-Universit\"at G\"ottingen, Institut f\"ur Astrophysik, Friedrich-Hund-Platz 1, D-37077 G\"ottingen, Germany}
\author{Ralf S.~Klessen}
 \email{klessen@uni-heidelberg.de}
 \affiliation{Universit\"at Heidelberg, Zentrum f\"ur Astronomie, Institut f\"ur Theoretische Astrophysik, Albert-\"Uberle-Strasse\ 2, D-69120 Heidelberg, Germany }

\date{\today}

\begin{abstract}
The present-day Universe is highly magnetized, even though the first magnetic seed fields were most probably extremely weak. To explain the growth of the magnetic field strength over many orders of magnitude fast amplification processes need to operate. The most efficient mechanism known today is the small-scale dynamo, which converts turbulent kinetic energy into magnetic energy leading to an exponential growth of the magnetic field. 
The efficiency of the dynamo depends on the type of turbulence indicated by the slope of the turbulence spectrum $v(\ell)\propto\ell^{\vartheta}$, where $v(\ell)$ is the eddy velocity at a scale $\ell$. We explore turbulent spectra ranging from incompressible Kolmogorov turbulence with $\vartheta=1/3$ to highly compressible Burgers turbulence with $\vartheta=1/2$. In this work we analyze the properties of the small-scale dynamo for low magnetic Prandtl numbers Pm, which denotes the ratio of the magnetic Reynolds number, Rm, to the hydrodynamical one, Re. We solve the Kazantsev equation, which describes the evolution of the small-scale magnetic field, using the WKB approximation. 
In the limit of low magnetic Prandtl numbers the growth rate is proportional to $\Rm^{(1-\vartheta)/(1+\vartheta)}$. We furthermore discuss the critical magnetic Reynolds number $\Rm_\text{crit}$, which is required for small-scale dynamo action. The value of $\Rm_\text{crit}$ is roughly 100 for Kolmogorov turbulence and 2700 for Burgers. Furthermore, we discuss that $\Rm_\text{crit}$ provides a stronger constraint in the limit of low Pm than it does for large Pm.
We conclude that the small-scale dynamo can operate in the regime of low magnetic Prandtl numbers, if the magnetic Reynolds number is large enough. Thus, the magnetic field amplification on small scales can take place in a broad range of physical environments and amplify week magnetic seed fields on short timescales.
\end{abstract}

\maketitle

\section{Introduction}

A large fraction of the Universe is magnetized. Various astrophysical phenomena have their origin in strong magnetic fields, for example jets from stars or galaxies and stellar activity. The question arises where those strong fields came from, especially because the generation mechanisms during inflation \citep{TurnerWidrow1988}, phase transitions in the early Universe \citep{Sigl1997} or battery processes \citep{Biermann1950,KulsrudEtAl1997,Xu2008} typically produce very week seed fields.\\
Magnetohydrodynamical dynamos are the most efficient mechanisms known to amplify weak magnetic seed fields. In particular the small-scale or turbulent dynamo is important as it converts turbulent kinetic energy into magnetic energy on very short timescales. \\
After the magnetic seed fields have been amplified exponentially by the kinematic dynamo on small scales, the nonlinear phase begins. The magnetic energy is transported through the inertial range up to the forcing scale of the turbulence on roughly the local eddy timescale. From this scale, which is about the Jeans scale in the case of a star forming region, the field can be transported to even larger scales by outflows of stars or supernovae.\\
How the small-scale dynamo operates in detail depends on the magnetic Prandtl number Pm, which is the ratio between kinematic viscosity $\nu$ and magnetic diffusivity $\eta$. With the hydrodynamic and magnetic Reynolds numbers $\Rey = VL/\nu$ and $\Rm = VL/\eta$, where $V$ is the typical velocity at the largest scale of the inertial range $L$, one can define $\mathrm{Pm} = \mathrm{Rm} / \mathrm{Re}$. While the small-scale dynamo analytically is well studied in the limit of infinite Pm \citep{Subramanian1997,SchekochihinEtAl2002,SchoberEtAl2012.1}, there are only a few studies for the case of $\Pm \rightarrow 0$ \citep{SchekochihinEtAl2005,SchekochihinEtAl2007,MalyshkinBoldyrev2010,KleeorinRogachevskii2012}. Up to now, simulations have been restricted to the regime of $0.1 \lesssim \Pm \lesssim 10$ \citep{FederrathEtAl2011.2}. \\
Nature features a broad range of magnetic Prandtl numbers, reaching from about $10^{12}$ in the primordial and present interstellar and intergalactic medium \citep{SchoberEtAl2012.2} to $10^{-7} - 10^{-2}$ in the interior of planets and stars \citep{SchekochihinEtAl2004,RobertsGlatzmaier2000}. The magnetic activity in the Sun is well explained by a large-scale dynamo model \citep{Parker1971,Parker1975, BrandenburgSubramanian2005}. However, there are observational indications that also the small-scale dynamo could play a role \citep{Miesch2012,CattaneoEtAl2003}. In addition, the small-scale dynamo could provide a source for large-scale dynamo action, especially if rotation is present \citep{PipinSeehafer2009}. Moreover, the regime of low magnetic Prandtl numbers is important in the case of liquid metal laboratory experiments, which also show the existence of a turbulent dynamo (see, e.g., \citet{NornbergEtAl2006}). These provide an additional comparison for theoretical results from the Kazantsev theory or simulations. \\
The efficiency of the small-scale dynamo depends on the type of turbulence. Most previous studies analyzed the case of ideal Kolmogorov turbulence \citep{Kolmogorov1941}, i.~e.~purely solenoidal turbulence. But astrophysical plasmas are usually highly compressible. For example, the energy and momentum input by supernova explosions lead to highly supersonic motions in the interstellar medium \cite{MacLowKlessen2004,ElmegreenScalo2004,ScaloElmegreen2004}. A similar interference is made for the accretion flow onto galactic disks or the convergent flows induced by spiral density waves \cite{HennebelleEtAl2008,BanerjeeEtAl2009,KlessenHennebelle2010}. Observations of the turbulent velocity spectrum confirm this argument \citep{Larson1981}. Here we take into account the effects of different types of turbulence \citep{SheLeveque1994,Boldyrev2002,Larson1981,Federrath2010,OssenkopfMacLow2002} ranging from incompressible Kolmogorov turbulence to highly compressible Burgers turbulence \citep{Burgers1948}. \\
In this paper we describe phenomenologically how the small-scale dynamo operates at $\Pm \rightarrow 0$ as motivated by the common stretch-twist-fold toy model, which was suggested for $\Pm \rightarrow \infty$. We summarize the concepts and the main equations of the Kazantsev theory, which analytically describes the small-scale dynamo in the kinematic limit. As input, one requires the correlation function of the turbulent velocity field (e.~g.~see the model presented in \citet{SchoberEtAl2012.1}). We employ the WKB approximation to solve the Kazantsev equation and find the growth rate of the magnetic field in the limit of small magnetic Prandtl numbers. The critical magnetic Reynolds number $\Rm_\text{crit}$, which needs to be exceeded for small-scale dynamo action, is the same for all magnetic Prandtl numbers analyzed. We discuss the influence of $\Rm_\text{crit}$ in the limit of small Pm. In the final section, we prove that the WKB approximation is valid for small to moderate Pm.

\section{Phenomenology of Small-Scale Dynamo Growth}

The small-scale dynamo converts kinetic energy from turbulent motions into magnetic energy. An illustrative model describing this process is the stretch-twist-fold dynamo \citep{VainshteinZeldovich1974}. The stretching of a closed magnetic flux rope leads to amplification of the magnetic field strength, as the magnetic flux is a conserved quantity. Afterwards the rope is stretched, twisted and folded such that the original shape is regained. The shorter the turnover time of the turbulent eddies is, the faster the stretch-twist-fold mechanism proceeds and thus the faster the magnetic field is amplified. Intuitively the turnover time decreases with decreasing eddy length. \\
In the limit of high magnetic Prandtl numbers the amplification rate of the dynamo is most efficient on the smallest scale of the inertial range, i.~e.~the viscous scale $\lnu = \Rey^{-1/(\vartheta+1)}~L$, where again $\vartheta$ is the slope of the turbulent velocity spectrum. \\
During the transition from large to small magnetic Prandtl numbers, the resistive scale $\leta = \Rm^{-1/(\vartheta+1)}~L$ becomes larger than the viscous one. The amplification then takes place at roughly $\leta$, which lies in the inertial range of the turbulent velocity spectrum. Due to larger time scales of the turbulent eddies in the inertial range, we expect the small-scale dynamo to be less efficient at low magnetic Prandtl numbers. While in the large Prandtl regime the hydrodynamical Reynolds number regulates the dynamo, here the magnetic Reynolds number is the relevant quantity.

\section{Analytical Description of the Small-Scale Dynamo}

\subsection{The turbulent velocity field}
A theoretical description of turbulence starts with the decomposition of the velocity field $\textbf{v}$ into a mean field $\left\langle\textbf{v}\right\rangle$ and a turbulent component $\delta\textbf{v}$:
\begin{equation}
  \textbf{v} = \left\langle\textbf{v}\right\rangle + \delta\textbf{v}.
\end{equation}
The correlation of two turbulent velocity components at the positions $\textbf{r}_1$ and $\textbf{r}_2$ at the times $t$ and $s$ for a Gaussian random velocity field with zero mean, which is isotropic, homogeneous, and $\delta$-correlated in time, is
\begin{equation}
  \left\langle \delta v_i(\textbf{r}_1,t)\delta v_j(\textbf{r}_2,s)\right\rangle = T_{ij}(r)\delta(t-s)
\end{equation}
with the two-point correlation function $T_{ij}(r)$ and $r\equiv|\textbf{r}_1-\textbf{r}_2|$. The delta-correlation in time is a simplifying assumption, and its consequences should be explored in future studies. Following \citet{Bachelor1953}, $T_{ij}(r)$ can be divided into a transverse part $T_{\text{N}}$ and a longitudinal part $T_{\text{L}}$ in the following way:
\begin{equation}
  T_{ij}(r) = \left(\delta_{ij}-\frac{r_i r_j}{r^2}\right)T_{\text{N}}(r) + \frac{r_ir_j}{r^2}T_{\text{L}}(r).
\end{equation}
We neglect here the effect of helicity, which would appear as an additional term in $T_{ij}$. Any turbulent flow can generally be described by the relation between the velocity $v(\ell)$ and the size $\ell$ of a velocity fluctuation,
\begin{equation}
  v(\ell) \propto \ell^{\vartheta}.
\label{TurbPower}
\end{equation}
The power-law index $\vartheta$ varies from its minimal value of $\vartheta=1/3$ for Kolmogorov theory \citep{Kolmogorov1941}, i.~e.~incompressible turbulence, to Burgers turbulence \citep{Burgers1948}, i.~e.~highly compressible turbulence, where $\vartheta$ gets its maximal value of $1/2$ \citep{Schmidt2009}. \\
We use the model for the correlation function of the turbulent velocity field from \citet{SchoberEtAl2012.1}. The longitudinal correlation function in the inertial range is motivated from the turbulent diffusion coefficient.  We ensure a continuous extension into the viscous range via an appropriate normalization. This leads to
\begin{equation}
  T_\text{L}(r) = \begin{cases} 
                     \frac{VL}{3}\left(1-\Rey^{(1-\vartheta)/(1+\vartheta)}\left(\frac{r}{L}\right)^{2}\right)        & 0<r<\ell_\nu \\ 
  		     \frac{VL}{3}\left(1-\left(\frac{r}{L}\right)^{\vartheta+1}\right)                                & \ell_\nu<r<L \\ 
  		     0                                                                                                & L<r, 
  		  \end{cases}
\label{TL}
\end{equation}
where $\ell_\nu=L~\Rey^{-1/(\vartheta+1)}$ denotes the cutoff scale of the turbulence, i.~e.~the viscous scale, and $L$ is the length of the largest eddies. In our model the transverse correlation function for the general slope of the turbulent velocity spectrum is
\begin{equation}
  T_\text{N}(r) = \begin{cases} 
                     \frac{VL}{3}\left(1-t(\vartheta)\Rey^{(1-\vartheta)/(1+\vartheta)} \left(\frac{r}{L}\right)^{2}\right)  	& 0<r<\ell_\nu \\
  		     \frac{VL}{3}\left(1-t(\vartheta)\left(\frac{r}{L}\right)^{\vartheta+1}\right)                              & \ell_\nu<r<L \\ 
  		     0                                                                                                   	& L<r, 
  	          \end{cases}
\label{TN}
\end{equation}
with $t(\vartheta)=(21-38\vartheta)/5$. The functional form of $T_\mathrm{N}$ is based on the relation between the transversal and longitudinal correlation function in the extreme cases of divergence-free (Kolmogorov) and rotation-free (Burgers) turbulence.

\subsection{Kazantsev Theory}
Like the velocity field, the magnetic field can be separated into a mean field $\left\langle\textbf{B}\right\rangle$ and a fluctuation part $\delta \textbf{B}$:
\begin{equation}
  \textbf{B} = \left\langle \textbf{B}\right\rangle + \delta \textbf{B}.
\end{equation}
Assuming that the fluctuating component $\delta \textbf{B}$ is a homogeneous, isotropic Gaussian random field with zero mean like the velocity field, we can write down the correlation function as
\begin{equation}
  \left\langle \delta B_i(\textbf{r}_1,t) \delta B_j(\textbf{r}_2,t)\right\rangle = M_{ij}(r,t)
\end{equation}
with the two-point correlation function
\begin{equation}
  M_{ij}(r,t) = \left(\delta_{ij}-\frac{r_ir_j}{r^2}\right)M_{\text{N}}(r,t) + \frac{r_ir_j}{r^2}M_{\text{L}}(r,t).
\label{correlationB}
\end{equation}
As the magnetic field is always divergence-free the transverse and the longitudinal correlation function are related by
\begin{equation}
  M_{\text{N}} = \frac{1}{2r} \frac{\text{d}}{\text{d} r}\left(r^2M_\text{L}\right),
\label{MN}
\end{equation}
where we have used that $(r_ir_j/r^2)M_{ij}=M_\text{L}$ and $(r_i/r_j)M_{ij}=M_\text{N}$. \\
The time derivative of $M_{ij}$ is
\begin{eqnarray}
  \frac{\partial M_{ij}}{\partial t} & = & \frac{\partial}{\partial t} \left\langle \delta B_i \delta B_j\right\rangle \nonumber \\
                                     & = & \left\langle\frac{\partial B_i}{\partial t}B_j\right\rangle + \left\langle B_i \frac{\partial B_j}{\partial t}\right\rangle \nonumber \\
                                     &   & -\frac{\partial}{\partial t}\left(\left\langle B_i\right\rangle\left\langle B_j\right\rangle\right).
\label{DeriMij}
\end{eqnarray}
In the upper equation we can substitute the induction equation
\begin{equation}
  \frac{\partial \textbf{B}}{\partial t} = \mathbf{\nabla}\times\textbf{v}\times\textbf{B} - \eta\mathbf{\nabla}\times\mathbf{\nabla}\times\textbf{B},
\label{induction}
\end{equation}
where $\eta \equiv c^2/(4\pi\sigma)$ is the magnetic diffusivity with the speed of light $c$ and the electrical conductivity $\sigma$, and the evolution equation of the magnetic mean field
\begin{equation}
  \frac{\partial\left\langle \textbf{B}\right\rangle}{\partial t} = \mathbf{\nabla} \times \left\langle \textbf{v}\right\rangle \times \left\langle \textbf{B}\right\rangle -\eta_\text{eff}\mathbf{\nabla} \times\mathbf{\nabla}\times\left\langle \textbf{B}\right\rangle
\label{meanB}
\end{equation}
with the effective parameter $\eta_\text{eff}=\eta+T_\text{L}(0)$. After a lengthy derivation \citep{BrandenburgSubramanian2005} this leads to
\begin{eqnarray}
  \frac{\partial M_\text{L}}{\partial t} & = & 2\kappa_\text{diff} M_\text{L}'' + 2\left(\frac{4\kappa_\text{diff}}{r}+ \kappa_\text{diff}'\right) M_\text{L}'    \nonumber \\
                                         &   & + \frac{4}{r}\left(\frac{T_\text{N}}{r} - \frac{T_\text{L}}{r} - T_\text{N}' - T_\text{L}'\right) M_\text{L} 
\label{dMLdt}
\end{eqnarray}
with
\begin{equation}
  \kappa_\text{diff}(r) = \eta + T_\text{L}(0) - T_\text{L}(r).
\label{kappaN}
\end{equation}
The prime denotes differentiation with respect to $r$. The diffusion of the magnetic correlations, $\kappa_\text{diff}$, contains in addition to the magnetic diffusivity $\eta$ the scale-dependent turbulent diffusion $T_\text{L}(0) - T_\text{L}(r)$.\\
With the solution of Eq.~(\ref{dMLdt}) we can calculate $M_\text{N}$ by using the relation (\ref{MN}). Thus, we find the total correlation function of the magnetic field fluctuations $M_{ij}$, which is proportional to the energy density of the fluctuating part of the magnetic field, $\delta B^2/(8\pi)$.\\
To separate the time from the spatial coordinates we use the ansatz 
\begin{equation}
  M_\text{L}(r,t) \equiv \frac{1}{r^2\sqrt{\kappa_\text{diff}}}\psi(r)\text{e}^{2\Gamma t}.
\end{equation}
Substitution of this ansatz in Eq.~ (\ref{dMLdt}) gives us
\begin{equation}
  -\kappa_\text{diff}(r)\frac{\text{d}^2\psi(r)}{\text{d}^2r} + U(r)\psi(r) = -\Gamma \psi(r).
\label{Kazantsev}
\end{equation}
This is the \textit{Kazantsev equation}, which is formally similar to the quantum-mechanical Schr\"odinger equation with a ``mass" $\hbar^2/(2\kappa_\text{diff})$ and the ``potential"\footnote{We note that there is a typo in the paper of \citet{SchoberEtAl2012.1}, where the potential was derived for a general type of turbulence. The term $2\kappa_\text{diff}$ / $r^2$ appeared here twice.}
\begin{equation}
  U(r) \equiv \frac{\kappa_\text{diff}''}{2} - \frac{(\kappa_\text{diff}')^2}{4\kappa_\text{diff}} + \frac{2\kappa_\text{diff}}{r^2} + \frac{2T_\text{N}'}{r} + \frac{2(T_\text{L}-T_\text{N})}{r^2}.
\label{GeneralPotential}
\end{equation}

\subsection{Formal Solution of the Kazantsev Equation in the WKB Approximation}
For the solution of the Kazantsev equation we use the WKB approximation. To use the standard formulation of this method, we have to make some substitutions. Definition of a new radial coordinate $x$ with $r \equiv \text{e}^x$ leads to
\begin{eqnarray}
  \frac{\kappa_\text{diff}(x)}{\text{e}^x}\frac{\text{d}}{\text{d}x}\left(\frac{1}{\text{e}^x}\frac{\text{d}\psi(x)}{\text{d}x}\right) - \left(\Gamma + U(x)\right)\psi(x)  =  0. 
\end{eqnarray}
Next we eliminate the first-derivative terms through the substitution
\begin{equation}
  \psi(x) \equiv \text{e}^{x/2}\theta(x),
\end{equation}
to obtain
\begin{equation}
  \frac{\text{d}^2\theta(x)}{\text{d}x^2} + p(x)\theta(x) = 0
\label{Kazantsev2}
\end{equation}
with the definition 
\begin{equation}
  p(x) \equiv  - \frac{[\Gamma + U(x)]\text{e}^{2x}}{\kappa_\text{diff}(x)} - \frac{1}{4}.
\label{p}
\end{equation}
The WKB solutions of Eq.~(\ref{Kazantsev2}) are linear combinations of
\begin{equation}
  \theta(x) = \frac{1}{p^{1/4}}\text{exp}\left(\pm i \int_{x_1}^{x} \sqrt{p(x')}\text{d}x'\right),
\label{eigenfunc}
\end{equation}
where $x_1$ is the first root of the $p$ function $p(x)$. The second derivative of (\ref{eigenfunc}) is 
\begin{equation}
  \theta''(x) + \left(1 + \frac{p''}{4p^2} - \frac{5}{16}\frac{(p')^2}{p^3}\right)p~\theta(x) = 0,                       
\end{equation}
where now the prime denotes $\text{d}/\text{d}x$. This equation results in the Kazantsev equation (\ref{Kazantsev}) if
\begin{equation}
  \left| f(x) \right| \ll 1,       
\label{CheckApprox}               
\end{equation}
with
\begin{equation}
  f(x) \equiv \frac{p''}{4p^2} - \frac{5}{16}\frac{(p')^2}{p^3}.     
\label{validityeq}  
\end{equation}
From the shape of the $p$-function we conclude that the solutions between the two roots of $p(x)$, i.~e.~$x_1<x<x_2$, are oscillatory. \\
The condition for the eigenvalues $\Gamma$ is \citep{MestelSubramanian1991}
\begin{equation}
  \int_{x_1}^{x_2}\sqrt{p(x')}\text{d}x' = \frac{2n+1}{2}\pi
\label{eigenvalue}
\end{equation}
for different excitation levels $n\in\mathbb{N}$. In this work we concentrate on the lowest mode $n=0$, which has the largest growth rate.

\section{Growth Rate in the Limit of Small Magnetic Prandtl Numbers}

\begin{figure}
  \includegraphics[width=0.48\textwidth]{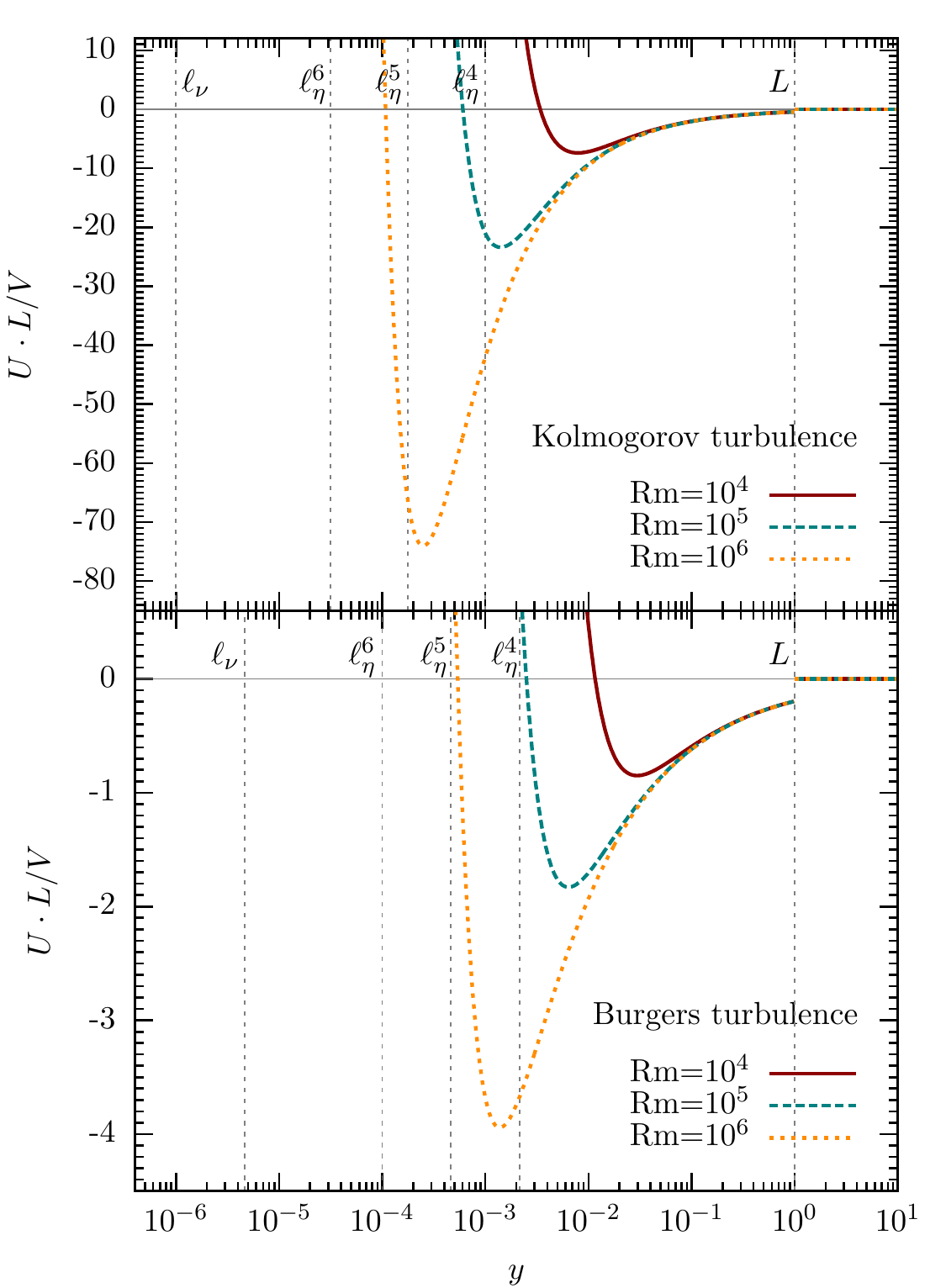}
  \caption{(Color online) Dependence of the potential on the dimensionless parameter $y\equiv r/L$ for Kolmogorov ($\vartheta=1/3$) and Burgers ($\vartheta=1/2$) turbulence at $\Rey=10^8$. We choose different magnetic Reynolds numbers $\Rm=10^4$, $\Rm=10^5$ and $\Rm=10^6$, resulting in the Prandtl numbers $\Pm=10^{-4}$, $\Pm=10^{-3}$ and $\Pm=10^{-2}$. The viscous scale $\ell_\nu$ depends on the type of turbulence and the Reynolds number. For Kolmogorov turbulence $\ell_\nu=\Rey^{-3/4}L$; for  Burgers turbulence $\ell_\nu=\Rey^{-2/3}L$. The resistive scale is  $\ell_\eta=\Rm^{-3/4}L$ for Kolmogorov and $\ell_\eta=\Rm^{-2/3}L$ for Burgers turbulence. A magnetic Reynolds number $10^x$ is indicated in the resistive scale as $\ell_\eta^\text{(x)}$. (Re appears only in the viscous range.)}
\label{plot_potential}
\end{figure}
\begin{figure}
  \includegraphics[width=0.48\textwidth]{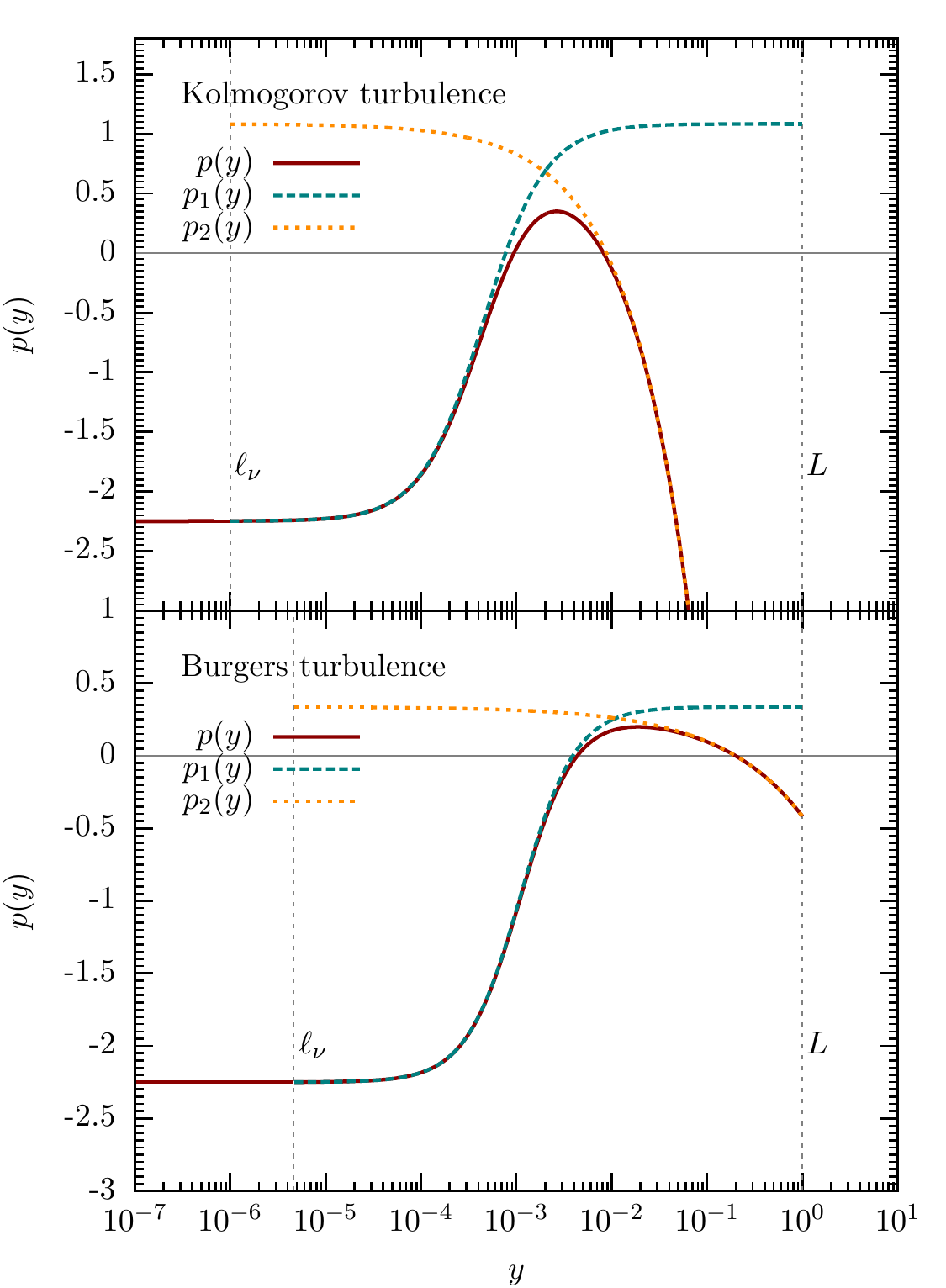}
  \caption{(Color online) The $p$ function (red curve) depending on the normalized scale parameter $y=r/l$ in the viscous, inertial and large-scale range. We indicate the viscous scale $\ell_\nu$ and the forcing scale $L$ as vertical lines. For the plot, we use our result for the growth rate in the limit of small Pm, $\Rey = 10^8$ and $\Rm = 10^{5}$. The dashed green line reefers to the approximation function $p_1$ and the dotted purple line to $p_2$. The upper panel shows $p(y)$ for Kolmogorov turbulence, the lower panel for Burgers turbulence.}
\label{pFunction_plot}
\end{figure}
\begin{figure}
  \includegraphics[width=0.48\textwidth]{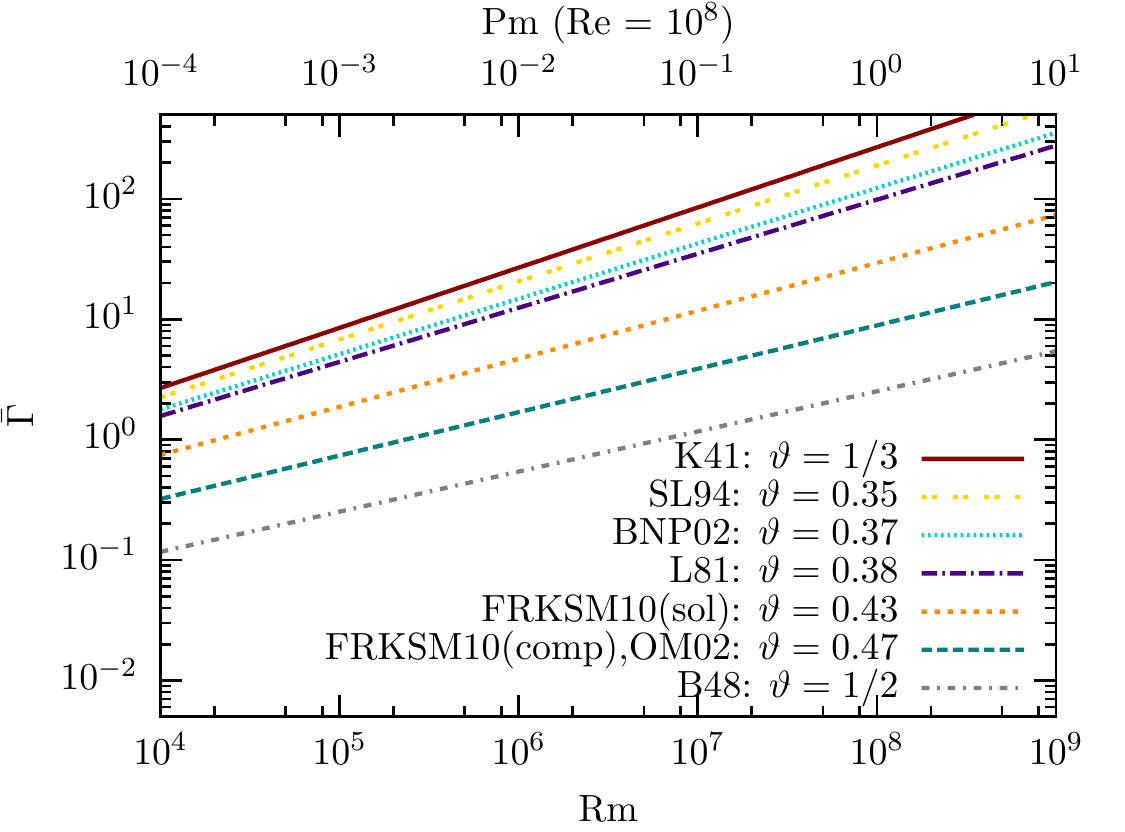}
  \caption{(Color online) The normalized growth rate $\bar{\Gamma}$ as a function of the magnetic Reynolds number Rm (lower x axes) and the magnetic Prandtl number Pm (upper x axes). The results shown for the lower abscissa are only valid for small Pm, i.~e., $\Rm \ll \Rey$, while we used a fixed Re of $10^{8}$ for the upper abscissa. We present different types for turbulence, indicated by the slope of the turbulent velocity spectrum $\vartheta$: K41 \cite{Kolmogorov1941}, SL94 \cite{SheLeveque1994}, BNP02 \cite{Boldyrev2002}, L81 \cite{Larson1981}, FRKSM10 \cite{Federrath2010} (sol: solenoidal forcing; comp: compressive forcing), OM02 \cite{OssenkopfMacLow2002} and B48 \cite{Burgers1948}.}
\label{Gamma_PmRm}
\end{figure}

We are interested in bound eigenfunctions of the Kazantsev equation (\ref{Kazantsev2}), which have corresponding real eigenvalues, i.~e.~growth rates. For this we require part of the potential (\ref{GeneralPotential}) to be negative. \\
In Fig.\ \ref{plot_potential} we show the normalized potential (\ref{GeneralPotential}) as a function of $y=r/L$ for Kolmogorov and Burgers turbulence. We choose a Reynolds number of $10^{8}$, which is a typical value for example for the interior of planets \cite{RobertsGlatzmaier2000} and primordial halos \citep{SchoberEtAl2012.2}. The different lines correspond to different magnetic Reynolds numbers of $10^{4}$, $10^{5}$ and $10^{6}$ and hence represent magnetic Prandtl numbers of $10^{-4}$, $10^{-3}$ and $10^{-2}$, respectively. The crucial discrepancy to the contrary limit of large $Pm$ is that the potential only has a negative part in the inertial range (i.~e.~the range between $\ell_\nu$ and $L$ indicated in the figure). Thus, there are only real positive eigenvalues of the Kazantsev equation (\ref{Kazantsev}) in this range.\\
With our model for the correlation function of the turbulent velocity field, Eqs.\ (\ref{TL}) and (\ref{TN}), the $p$ function (\ref{p}) in the inertial range is 
\begin{eqnarray}
  p(y) & = & \frac{-3}{20\,{\left( 3 + \Rm\,y^{1 + \vartheta}\right)}^2}\,\left(135 + \Rm\,y\,\left(60\,y\,\bar{\Gamma}\right.\right. \nonumber \\
       &   & \left.\left. -a(\vartheta)\,\Rm\,y^{1 + 2\,\vartheta}  \right.\right. \nonumber \\
       &   & \left.\left. + 2\,y^{\vartheta }\,\left( 25 - b(\vartheta) + 10\,\Rm\,y^2\,\bar{\Gamma}\right)\right)\right), \nonumber \\
\label{p2}
\end{eqnarray}
where 
\begin{eqnarray}
  \bar{\Gamma} = \frac{L}{V}~\Gamma 
\end{eqnarray}
is the normalized growth rate, and we use the abbreviations 
\begin{eqnarray}
  a(\vartheta) & = & \vartheta(56-103\vartheta) \\
  b(\vartheta) & = & \vartheta(79-157\vartheta).
\end{eqnarray}
For the analytical determination of the zeros of $p(y)$ we use the approximations
\begin{equation}
  p_1(y) = \frac{3\,\Rm\,y^{1 + \vartheta }\,\left(a(\vartheta)\,\Rm\,y^{1 + \vartheta } + 2\,b(\vartheta) - 50\right)-405}{20\,{\left( 3 + \Rm\,y^{1 + \vartheta } \right) }^2},
\end{equation}
which is valid for $\bar{\Gamma}\rightarrow 0$, and
\begin{equation}
  p_2(y) = \frac{3\,a(\vartheta)}{20} - 3\,y^{1 - \vartheta }\,\bar{\Gamma},
\end{equation}
where we leave out the constant terms in (\ref{p2}). We show $p(y)$ as well as the two approximations in Fig.\ \ref{pFunction_plot} for the exemplary case of $\Rey=10^8$ and $\Rm = 10^{5}$.\\
By using $p_1(y)$ we find for the first zero of $p(y)$ approximately
\begin{equation}
  y_1 = {\left( \frac{c(\vartheta)}{\Rm} \right) }^{\frac{1}{1 + \vartheta}},
\end{equation}
where we defined
\begin{equation}
  c(\vartheta) = \frac{25 + {\sqrt{135\,a(\vartheta) + {\left(b(\vartheta) - 25\right)}^2}} - b(\vartheta)}{a(\vartheta)}.
\end{equation}
With $p_2(y)$ we find the second zero
\begin{equation}
  y_2 = \left(\frac{a(\vartheta)}{20\,\bar{\Gamma}}\right)^{\frac{1}{1 - \vartheta}}.
\end{equation}
The eigenvalue can be determined approximately by the equation
\begin{equation}
  \int_{y_1}^{y_2} \frac{\sqrt{p_2(y)}}{y} \text{d}y = \frac{\pi}{2}.
\label{eigenvalue_y}
\end{equation}
Here we use the approximative function $p_2(y)$ instead of the full function $p(y)$ in order to find an analytical solution of the integral. Note that the scaling of the abscissa in Fig.\ \ref{pFunction_plot} is logarithmic and thus $p_2(y)$ is a good approximation of $p(y)$ for $y > y_1$. The value of the integral does not change by much due to this simplification.\\
We can solve the resulting equation from (\ref{eigenvalue_y}) with the ansatz
\begin{equation}
  \bar{\Gamma} = \alpha~\Rm^{\frac{1-\vartheta}{1+\vartheta}}.
\label{ansatz}
\end{equation}
This is motivated by the result of \citet{SchoberEtAl2012.1} in the limit of large magnetic Prandtl numbers: $\bar{\Gamma} \propto Re^{(1-\vartheta)/(1+\vartheta)}$. Here the amplification process takes place at the viscous scale, which depends on the hydrodynamical Reynolds number. As mentioned above in the limit of low magnetic Prandtl numbers the dynamo operates mainly on the resistive scale, which depends on Rm. Thus, our ansatz is to replace the hydrodynamic Reynolds number by the magnetic one (see also e.g.~\citet{BoldyrevCattaneo2004}).\\
With (\ref{ansatz}) we find for the solution of (\ref{eigenvalue_y}):
\begin{eqnarray}
  \frac{1} {\vartheta - 1}\,\sqrt{\frac{3}{5}}\,\left( {\sqrt{a(\vartheta) - 20\,c(\vartheta)^{\frac{1 - \vartheta }{1 + \vartheta }}\,\alpha }} + \right. & & \nonumber \\
  \left. \sqrt{a(\vartheta)}\,\log\Big(4\,\sqrt{5}\,\sqrt{{\Rm}^{\frac{1 - \vartheta }{1 + \vartheta }}\,\alpha}\Big) - \right. & & \nonumber \\
  \left. \sqrt{a(\vartheta)}\,\log\Big(2\,{\left(\frac{c(\vartheta)}{\Rm} \right) }^{\frac{\vartheta - 1}{2\,\left(1 + \vartheta\right) }}\,\left(\sqrt{a(\vartheta)} \right. \right. & & \nonumber \\
  \left.\left. + \sqrt{a(\vartheta) - 20\,c(\vartheta)^{\frac{1 - \vartheta }{1 + \vartheta }}\,\alpha}\right)\Big)\right) & = & \frac{\pi}{2}.
\label{eigenvalue_eq}
\end{eqnarray}
As we assume the pre-factor of the growth rate to be very small, i.~e.~$\alpha\ll1$, we use $a(\vartheta) \gg 20\,c(\vartheta)^{(1 - \vartheta)(1 + \vartheta)}\,\alpha$ to approximate (\ref{eigenvalue_eq}) as
\begin{eqnarray}
  \frac{{\sqrt{a(\vartheta)}}}{\vartheta - 1}\,\sqrt{\frac{3}{5}}\,\left( 1 - \log\Big(4\,{\sqrt{a(\vartheta)}}\,{\left( \frac{c(\vartheta)}{\Rm} \right) }^
  {\frac{\vartheta - 1}{2\,\left(1 + \vartheta\right)}}\Big)\right. & & \nonumber \\
  \left. + \log\Big(4\,{\sqrt{5}}\,{\sqrt{{\Rm}^{\frac{1 - \vartheta }{1 + \vartheta }}\,\alpha }}\Big)\right) & = & \frac{\pi}{2}. \nonumber \\
  \label{eigenvalue_eq_approx}
\end{eqnarray}
The solution of this equation can easily be found:
\begin{equation}
  \alpha =\frac{a(\vartheta)}{5}\,c(\vartheta)^{\frac{\vartheta - 1}{1 + \vartheta }}\,\exp\left(\sqrt{\frac{5}{3\,a(\vartheta)}}\,\pi \,\left(\vartheta - 1\right) - 2\right).
\end{equation}
We list results for the normalized growth rate of the small-scale dynamo in the limit of low magnetic Prandtl numbers for exemplary types of turbulence in Tab.~\ref{ResultsTable}. For comparison we also list the results for $\Pm\rightarrow\infty$ from \citet{SchoberEtAl2012.1}. Moreover, we present the normalized growth rate in Fig.\ \ref{Gamma_PmRm}. The lower abscissa shows the dependency on the magnetic Reynolds number, which is valid for any hydrodynamic Reynolds number $\Rey \gg \Rm$. The upper abscissa in Fig.\ \ref{Gamma_PmRm} shows the dependency on the magnetic Prandtl number for a fixed Re of $10^{20}$. We present the results for different types of turbulence reported in the astrophysical literature \citep{Kolmogorov1941,SheLeveque1994,Boldyrev2002,Larson1981,Federrath2010,OssenkopfMacLow2002,Burgers1948}.

\begin{table*}    
  \begin{ruledtabular}
    \begin{tabular}{lcccc}
      \parbox[0pt][2.5em][c]{0cm}{}  Model and reference                     &  $\vartheta$    &  $\bar{\Gamma}$ (Rm$\ll$Re) &  $\bar{\Gamma}$ (Rm$\gg$Re)   &  $\Rm_\text{crit}$    \\
      \hline
      \tabhe \citet{Kolmogorov1941}                                          &  $1/3$          &  $0.027~\Rm^{1/2}$   &  $1.03~\Rey^{1/2}$      &  $\approx 107$              \\
      \tabhe Intermittency of Kolmogorov turbulence \citep{SheLeveque1994}   &  $0.35$         &  $0.027~\Rm^{0.48}$  &  $0.94~\Rey^{0.48}$     &  $\approx 118$              \\
      \tabhe Driven supersonic MHD-turbulence \citep{Boldyrev2002}           &  $0.37$         &  $0.026~\Rm^{0.46}$  &  $0.84~\Rey^{0.46}$     &  $\approx 137$              \\
      \tabhe Observation in molecular clouds \citep{Larson1981}              &  $0.38$         &  $0.025~\Rm^{0.45}$  &  $0.79~\Rey^{0.45}$     &  $\approx 149$              \\
      \tabhe Solenoidal forcing of the turbulence \citep{Federrath2010}      &  $0.43$         &  $0.019~\Rm^{0.40}$  &  $0.54~\Rey^{0.40}$     &  $\approx 227$              \\
      \tabhe Compressive forcing of the turbulence \citep{Federrath2010}     &  $0.47$         &  $0.012~\Rm^{0.36}$  &  $0.34~\Rey^{0.36}$     &  $\approx 697$              \\
      \tabhe Observations in molecular clouds \citep{OssenkopfMacLow2002}    &  $0.47$         &  $0.012~\Rm^{0.36}$  &  $0.34~\Rey^{0.36}$     &  $\approx 697$              \\
      \tabhe \citet{Burgers1948}                                             &  $1/2$          &  $0.0054~\Rm^{1/3}$  &  $0.18~\Rey^{1/3}$      &  $\approx 2718$             \\
    \end{tabular}
  \end{ruledtabular}
\caption{The normalized growth rate of the small-scale dynamo $\bar{\Gamma}$ in the limit of small magnetic Prandtl numbers (Rm$\ll$Re). For comparison, we present also $\bar{\Gamma}$ for large magnetic Prandtl numbers (Rm$\gg$Re). We show our results for different types of turbulence, which are characterized by the exponent $\vartheta$ of the slope of the turbulent velocity spectrum, $v(\ell)\propto\ell^\vartheta$. The extreme values of $\vartheta$ are $1/3$ for Kolmogorov turbulence and $1/2$ for Burgers turbulence.}
  \label{ResultsTable}
\end{table*}

\section{Critical Magnetic Reynolds Number}

For the onset of the small-scale dynamo the magnetic Reynolds number needs to exceed a critical value $\Rm_\text{crit}$. We determine the latter by setting the growth rate in (\ref{p}) equal to zero and solving Eq.~(\ref{eigenvalue}) for Rm. As the $p$-function in the inertial range only depends on Rm, but not on Re, it is independent of the magnetic Prandtl number. We list the numerical results for $\Rm_\text{crit}$ in Tab.\ \ref{ResultsTable}. The critical magnetic Reynolds number increases with increasing compressibility.\\
In the limit of large Pm the critical magnetic Reynolds number is not necessarily the dominant restriction, as $\mathrm{Re} > 10^3$ is required for turbulent flows. As $\Rm \gg \Rey$ for large Pm, Rm needs to be much larger than $10^3$, which is larger than the critical magnetic Reynolds number.\\
In the opposite limit of small Pm we have the case of $\Rm \ll \Rey$. For low hydrodynamic Reynolds numbers, the magnetic Reynolds number can fall below $\Rm_\text{crit}$, and the small-scale dynamo can not operate.\\
We note that, contrary to our results presented here, \citet{IskakovEtAl2007} found a weak dependence of the critical magnetic Reynolds number on the magnetic Prandtl number. Therefore, it needs to be explored further whether the discrepancy in our results is due to approximations in the Kazantsev model or if it is a result of the relatively narrow inertial range in numerical simulations. The latter provides a restriction on the number of turbulent eddies resolved in the box, and thus on the overall statistical sampling of the dynamics. In future studies, it would thus be desirable to explore this behavior at higher resolution in numerical simulations, and by relaxing the assumption of the Kazantsev theory in analytical studies.

\section{Comparison with Numerical Solution}

\begin{figure}
  \includegraphics[width=0.48\textwidth]{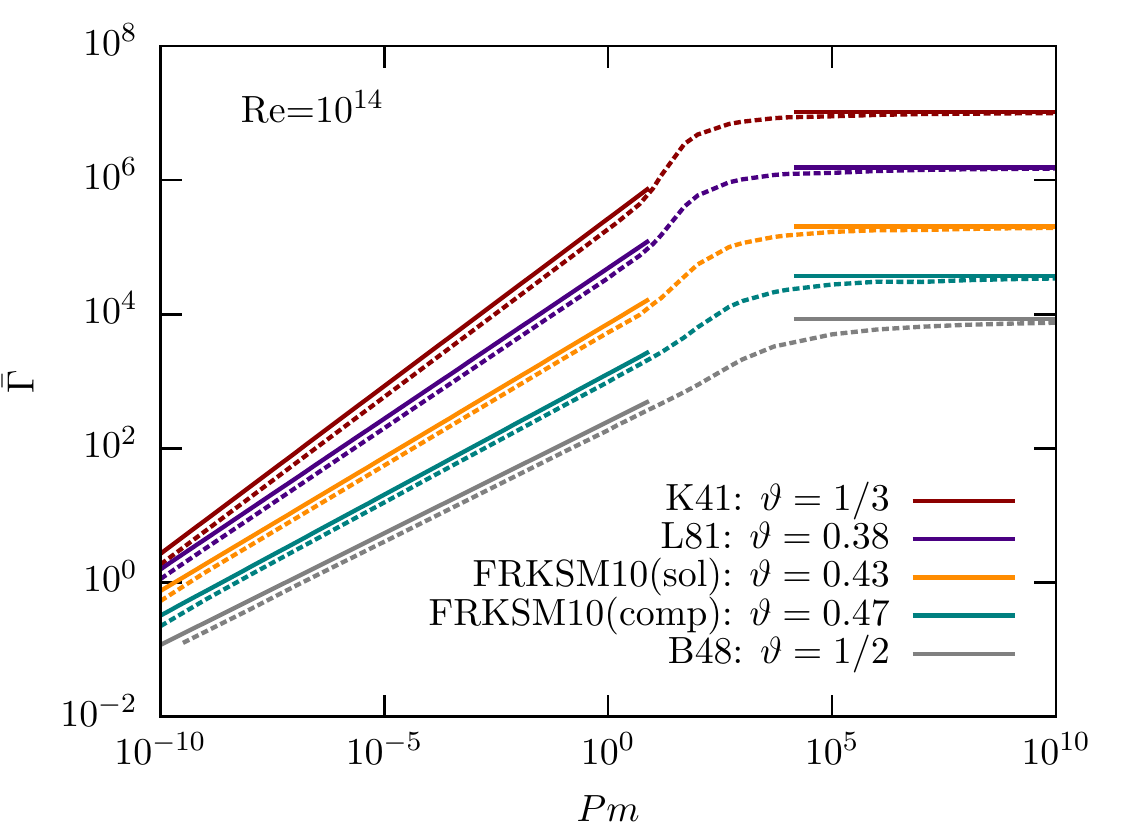}
  \caption{(Color online) The normalized growth rate $\bar{\Gamma}$ as a function of the magnetic Prandtl number Pm. The results are shown for a fixed Reynolds number $\Rey = 10^{14}$. We present the solutions from numerical integration of the Kazantsev equation by \citet{BovinoSchleicherSchober2012} indicated as dashed lines. The analytical solutions in the limits of small, i.~e., Eq.~(\ref{ansatz}), and large Prandtl numbers \cite{SchoberEtAl2012.1} are shown by the solid lines. We present different types for turbulence, indicated by the slope of the turbulent velocity spectrum $\vartheta$: K41 \cite{Kolmogorov1941}, L81 \cite{Larson1981}, FRKSM10 \cite{Federrath2010} (sol: solenoidal forcing; comp: compressive forcing), and B48 \cite{Burgers1948}.}
\label{Gamma_AnaNum}
\end{figure}

\citet{BovinoSchleicherSchober2012} solved the Kazantsev equation (\ref{Kazantsev}) numerically with the Numerov algorithm. They used the same form of the correlation functions of the turbulent velocity, i.~e.~ Eqs.~(\ref{TL}) and (\ref{TN}). A comparison of their result with our analytical solution is shown in Fig.~\ref{Gamma_AnaNum} at a fixed Reynolds number of $10^{14}$.\\
The comparison shows excellent agreement between the numerical and the analytical solutions in the limit of small ($\bar{\Gamma} \propto \Rm^{(1-\vartheta)/(1+\vartheta)}$) and large magnetic Prandtl numbers ($\bar{\Gamma} \propto \Rey^{(1-\vartheta)/(1+\vartheta)}$ \cite{SchoberEtAl2012.1}). We find that the range where our analytical solution (\ref{ansatz}) can be used is not restricted to $\Pm \ll 1$ but is also applicable in the regime $\Pm \approx 1$ for all types of turbulence. We see a minor offset between our solutions and the numerical ones for small Pm. This is probably caused by the approximations we made from Eqs.~(\ref{eigenvalue_eq}) and (\ref{eigenvalue_eq_approx}).

\section{Conclusion}
In this paper, we used the Kazantsev theory to determine the growth rate of the small-scale dynamo in the limit of low to moderate magnetic Prandtl numbers. We found that the growth rate is proportional to $\Rm^{(1-\vartheta)/(1+\vartheta)}$, where Rm is the magnetic Reynolds number and $\vartheta$ is the slope of the turbulent velocity spectrum in the inertial range. The critical magnetic Reynolds number for small-scale dynamo action $\Rm_\text{crit}$ ranges from roughly 100 for Kolmogorov turbulence ($\vartheta=1/3$) to 2700 for Burgers turbulence ($\vartheta=1/2$). These values are the same for large and low magnetic Prandtl numbers. However, for large Pm $\Rm_\text{crit}$ provides no strong constraint, as here $\Rm\gg\Rey$ and $\Rey\gtrsim 10^3$ for turbulence. In the limit of small Pm, where $\Rm\ll\Rey$, the critical magnetic Reynolds number is more important. We derived our results employing the WKB approximation, which we have shown to accurately solve the Kazantsev equation in the limit of small magnetic Prandtl numbers (see Appendix \ref{Validity}). Numerical integration of the Kazantsev equation predicts a smooth transition of the growth rate between the small and large Prandtl number regime (see Fig. \ref{Gamma_AnaNum} and \cite{BovinoSchleicherSchober2012}). The analytical solutions for $\Pm\ll 1$ and $\Pm\gg 1$ cover a broad range of possible magnetic Prandtl numbers and are also applicable at $\Pm\approx 1$. This helps us to better understand numerical simulations, which, because of the limited resolution that can be achieved, are bound to magnetic Prandtl numbers around unity.\\
We gained important results for the small-scale dynamo from the Kazantsev theory, which are summarized in Table \ref{ResultsTable}. However, one has to be careful with the indicated numerical values as these show only the expected trends resulting from our assumptions. The growth rates as well as the critical magnetic Reynolds numbers can change if we take additional physical mechanisms into account, such as helicity or the finite correlation time of the turbulent eddies. Future highly resolved numerical simulations will provide a basis for comparison and help to determine how strongly these additional effects may change the properties of the turbulent dynamo. \\
With our calculations we show that small-scale dynamo action is not restricted to the regime of large magnetic Prandtl numbers, but it also occurs at low and moderate Pm. This can be used to analyze the evolution of the small-scale magnetic field in various physical environments in more detail. Rapid amplification of magnetic seed fields is necessary to explain the strength of magnetic fields in the present-day Universe.

\acknowledgements
We are grateful to Christoph Federrath, Robi Banerjee and Wolfram Schmidt for useful discussions and to the anonymous referees for valuable suggestions on the manuscript. We thank for funding through the {\em Deutsche Forschungsgemeinschaft} (DFG) in the {\em Schwer\-punkt\-programm} SPP 1573 ``Physics of the Interstellar Medium" under grant KL 1358/14-1 and SCHL 1964/1-1. Moreover, we thank for financial support by the {\em Baden-W\"urttemberg-Stiftung} via contract research (grant P-LS-SPII/18) in their program ``Internationale Spitzenforschung II" as well as the DFG via the SFB 881 ``The Milky Way System" in the sub-projects B1 and B2. J.~S. acknowledges the support by IMPRS HD. D.~R.~G.~S.~thanks for funding via the SFB 963/1 on ``Astrophysical flow instabilities and turbulence".

\appendix

\section{Validity of the WKB Approximation}
\label{Validity}

\begin{figure}
  \includegraphics[width=0.48\textwidth]{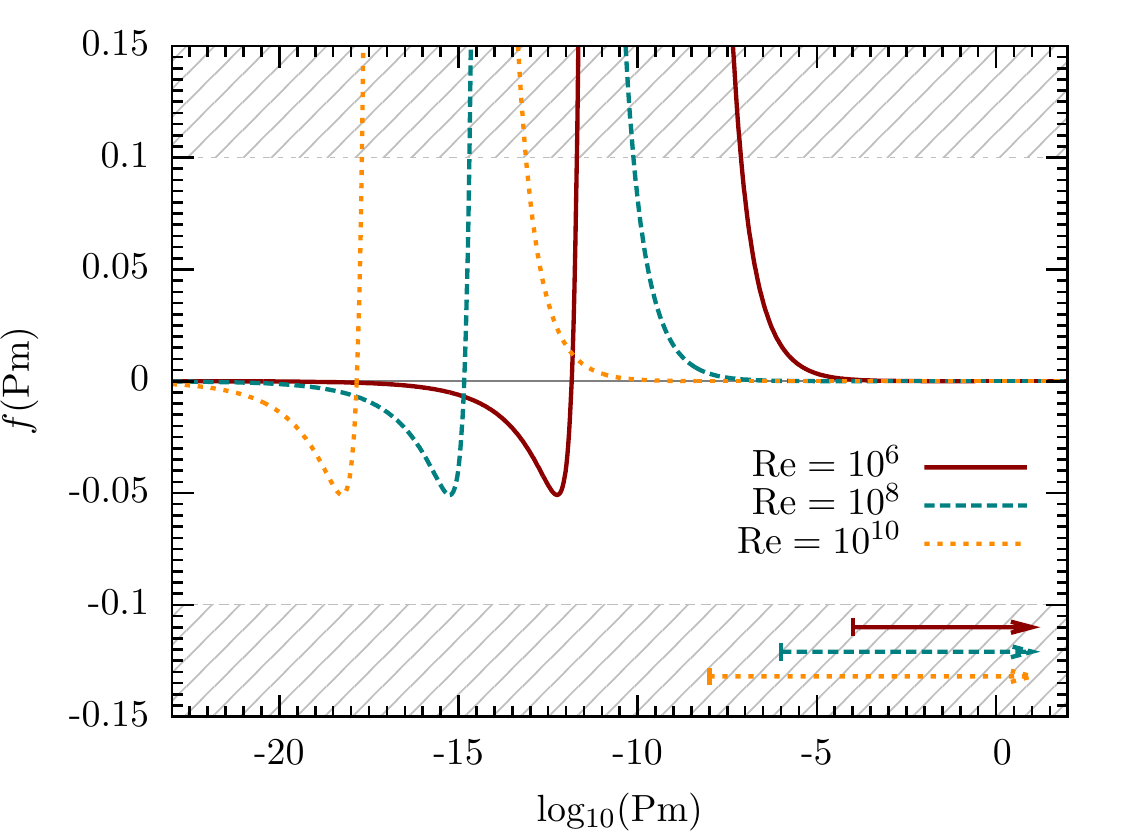}
  \caption{(Color online) Test of the validity of the WKB approximation for Kolmogorov turbulence. The function (\ref{validityeq}) is shown depending on the magnetic Prandtl number Pm for different hydrodynamic Reynolds numbers $\Rey=10^6$, $\Rey=10^8$ and $\Rey=10^{10}$. We evaluated $f(\Pm)$ at the minimum of the potential (\ref{GeneralPotential}). The WKB approximation is valid for $f(\Pm) \rightarrow 0$. The critical magnetic Reynolds number gives a further restriction (see text) leading to possible values for Pm of $\Pm \gtrsim 10^{-4}$ ($\Rey=10^6$), $\Pm \gtrsim 10^{-6}$ ($\Rey=10^8$) and $\Pm \gtrsim 10^{-8}$ ($\Rey=10^{10}$), which we indicated by the corresponding arrows.}
\label{Validity_Kolmo}
\end{figure}

In this section we show that the WKB approximation, which we use for solving the Kazantsev equation, is valid in the limit of small magnetic Prandtl numbers. Therefore, we have to analyze (\ref{validityeq}) for the inertial range of the turbulence spectrum. As $f$ is a function of the distance $x$, i.~e.~$r$, we have to evaluate it on the characteristic scale in which we are interested. With the main amplification occurring at the minimum of the potential, it is rational to use this scale. \\
In Fig.\ \ref{Validity_Kolmo} we show $f(\Pm)$ on the scale of the potential minimum for different Reynolds numbers, $\Rey=10^6$, $\Rey=10^{8}$ and $\Rey=10^{10}$. We choose the example of Kolmogorov turbulence for the discussion. The test of validity is of course similar for other types of turbulence.\\
The magnetic Reynolds number needs to exceed $\Rm_\text{crit}$: 
\begin{equation}
  \Rm = \Pm~\Rey > \Rm_\text{crit}.
\end{equation}
For Kolmogorov turbulence $\Rm_\text{crit} \approx 10^2$. Thus, for example in the curve with $\Rey=10^{8}$ in Fig.\ \ref{Validity_Kolmo} the threshold is only exceeded for $\Pm > 10^{-6}$. In this regime the WKB approximation is perfectly valid. In Fig.\ \ref{Validity_Kolmo} we use arrows to indicate the regimes, where the small-scale dynamo can operate.\\
In principle, Fig.\ \ref{Validity_Kolmo} states that our approximation is valid also for larger Pm up to $\Pm \rightarrow \infty$. However, at some point we face the problem in which the potential gets negative also below the viscous range. For $\Pm \rightarrow \infty$, the negative part of the potential in the viscous range clearly dominates. As we do not account for this range in the calculations above, our results are only valid for sufficiently small Pm. 


\begin{thebibliography}{44}%
\makeatletter
\providecommand \@ifxundefined [1]{%
 \@ifx{#1\undefined}
}%
\providecommand \@ifnum [1]{%
 \ifnum #1\expandafter \@firstoftwo
 \else \expandafter \@secondoftwo
 \fi
}%
\providecommand \@ifx [1]{%
 \ifx #1\expandafter \@firstoftwo
 \else \expandafter \@secondoftwo
 \fi
}%
\providecommand \natexlab [1]{#1}%
\providecommand \enquote  [1]{``#1''}%
\providecommand \bibnamefont  [1]{#1}%
\providecommand \bibfnamefont [1]{#1}%
\providecommand \citenamefont [1]{#1}%
\providecommand \href@noop [0]{\@secondoftwo}%
\providecommand \href [0]{\begingroup \@sanitize@url \@href}%
\providecommand \@href[1]{\@@startlink{#1}\@@href}%
\providecommand \@@href[1]{\endgroup#1\@@endlink}%
\providecommand \@sanitize@url [0]{\catcode `\\12\catcode `\$12\catcode
  `\&12\catcode `\#12\catcode `\^12\catcode `\_12\catcode `\%12\relax}%
\providecommand \@@startlink[1]{}%
\providecommand \@@endlink[0]{}%
\providecommand \url  [0]{\begingroup\@sanitize@url \@url }%
\providecommand \@url [1]{\endgroup\@href {#1}{\urlprefix }}%
\providecommand \urlprefix  [0]{URL }%
\providecommand \Eprint [0]{\href }%
\providecommand \doibase [0]{http://dx.doi.org/}%
\providecommand \selectlanguage [0]{\@gobble}%
\providecommand \bibinfo  [0]{\@secondoftwo}%
\providecommand \bibfield  [0]{\@secondoftwo}%
\providecommand \translation [1]{[#1]}%
\providecommand \BibitemOpen [0]{}%
\providecommand \bibitemStop [0]{}%
\providecommand \bibitemNoStop [0]{.\EOS\space}%
\providecommand \EOS [0]{\spacefactor3000\relax}%
\providecommand \BibitemShut  [1]{\csname bibitem#1\endcsname}%
\let\auto@bib@innerbib\@empty
\bibitem [{\citenamefont {Turner}\ and\ \citenamefont
  {Widrow}(1988)}]{TurnerWidrow1988}%
  \BibitemOpen
  \bibfield  {author} {\bibinfo {author} {\bibfnamefont {M.~S.}\ \bibnamefont
  {Turner}}\ and\ \bibinfo {author} {\bibfnamefont {L.~M.}\ \bibnamefont
  {Widrow}},\ }\href {\doibase 10.1103/PhysRevD.37.2743} {\bibfield  {journal}
  {\bibinfo  {journal} {Phys. Rev. D}\ }\textbf {\bibinfo {volume} {37}},\
  \bibinfo {pages} {2743} (\bibinfo {year} {1988})}\BibitemShut {NoStop}%
\bibitem [{\citenamefont {{Sigl}}, \citenamefont {{Olinto}},\ and\
  \citenamefont {{Jedamzik}}(1997)}]{Sigl1997}%
  \BibitemOpen
  \bibfield  {author} {\bibinfo {author} {\bibfnamefont {G.}~\bibnamefont
  {{Sigl}}}, \bibinfo {author} {\bibfnamefont {A.~V.}\ \bibnamefont
  {{Olinto}}}, \ and\ \bibinfo {author} {\bibfnamefont {K.}~\bibnamefont
  {{Jedamzik}}},\ }\href {\doibase 10.1103/PhysRevD.55.4582} {\bibfield
  {journal} {\bibinfo  {journal} {\prd}\ }\textbf {\bibinfo {volume} {55}},\
  \bibinfo {pages} {4582} (\bibinfo {year} {1997})}\BibitemShut {NoStop}%
\bibitem [{\citenamefont {{Biermann}}(1950)}]{Biermann1950}%
  \BibitemOpen
  \bibfield  {author} {\bibinfo {author} {\bibfnamefont {L.}~\bibnamefont
  {{Biermann}}},\ }\href@noop {} {\bibfield  {journal} {\bibinfo  {journal}
  {Zeitschrift Naturforschung Teil A}\ }\textbf {\bibinfo {volume} {5}},\
  \bibinfo {pages} {65} (\bibinfo {year} {1950})}\BibitemShut {NoStop}%
\bibitem [{\citenamefont {{Kulsrud}}\ \emph {et~al.}(1997)\citenamefont
  {{Kulsrud}}, \citenamefont {{Cowley}}, \citenamefont {{Gruzinov}},\ and\
  \citenamefont {{Sudan}}}]{KulsrudEtAl1997}%
  \BibitemOpen
  \bibfield  {author} {\bibinfo {author} {\bibfnamefont {R.}~\bibnamefont
  {{Kulsrud}}}, \bibinfo {author} {\bibfnamefont {S.~C.}\ \bibnamefont
  {{Cowley}}}, \bibinfo {author} {\bibfnamefont {A.~V.}\ \bibnamefont
  {{Gruzinov}}}, \ and\ \bibinfo {author} {\bibfnamefont {R.~N.}\ \bibnamefont
  {{Sudan}}},\ }\href {\doibase 10.1016/S0370-1573(96)00061-0} {\bibfield
  {journal} {\bibinfo  {journal} {\physrep}\ }\textbf {\bibinfo {volume}
  {283}},\ \bibinfo {pages} {213} (\bibinfo {year} {1997})}\BibitemShut
  {NoStop}%
\bibitem [{\citenamefont {{Xu}}\ \emph {et~al.}(2008)\citenamefont {{Xu}},
  \citenamefont {{O'Shea}}, \citenamefont {{Collins}}, \citenamefont
  {{Norman}}, \citenamefont {{Li}},\ and\ \citenamefont {{Li}}}]{Xu2008}%
  \BibitemOpen
  \bibfield  {author} {\bibinfo {author} {\bibfnamefont {H.}~\bibnamefont
  {{Xu}}}, \bibinfo {author} {\bibfnamefont {B.~W.}\ \bibnamefont {{O'Shea}}},
  \bibinfo {author} {\bibfnamefont {D.~C.}\ \bibnamefont {{Collins}}}, \bibinfo
  {author} {\bibfnamefont {M.~L.}\ \bibnamefont {{Norman}}}, \bibinfo {author}
  {\bibfnamefont {H.}~\bibnamefont {{Li}}}, \ and\ \bibinfo {author}
  {\bibfnamefont {S.}~\bibnamefont {{Li}}},\ }\href {\doibase 10.1086/595617}
  {\bibfield  {journal} {\bibinfo  {journal} {\apjl}\ }\textbf {\bibinfo
  {volume} {688}},\ \bibinfo {pages} {L57} (\bibinfo {year}
  {2008})}\BibitemShut {NoStop}%
\bibitem [{\citenamefont {{Subramanian}}(1997)}]{Subramanian1997}%
  \BibitemOpen
  \bibfield  {author} {\bibinfo {author} {\bibfnamefont {K.}~\bibnamefont
  {{Subramanian}}},\ }\href@noop {} {\bibfield  {journal} {\bibinfo  {journal}
  {ArXiv e-prints}\ } (\bibinfo {year} {1997})},\ \bibinfo {note}
  {arXiv:astro-ph/9708216}\BibitemShut {NoStop}%
\bibitem [{\citenamefont {{Schekochihin}}\ \emph {et~al.}(2002)\citenamefont
  {{Schekochihin}}, \citenamefont {{Cowley}}, \citenamefont {{Hammett}},
  \citenamefont {{Maron}},\ and\ \citenamefont
  {{McWilliams}}}]{SchekochihinEtAl2002}%
  \BibitemOpen
  \bibfield  {author} {\bibinfo {author} {\bibfnamefont {A.~A.}\ \bibnamefont
  {{Schekochihin}}}, \bibinfo {author} {\bibfnamefont {S.~C.}\ \bibnamefont
  {{Cowley}}}, \bibinfo {author} {\bibfnamefont {G.~W.}\ \bibnamefont
  {{Hammett}}}, \bibinfo {author} {\bibfnamefont {J.~L.}\ \bibnamefont
  {{Maron}}}, \ and\ \bibinfo {author} {\bibfnamefont {J.~C.}\ \bibnamefont
  {{McWilliams}}},\ }\href {\doibase 10.1088/1367-2630/4/1/384} {\bibfield
  {journal} {\bibinfo  {journal} {New Journal of Physics}\ }\textbf {\bibinfo
  {volume} {4}},\ \bibinfo {pages} {84} (\bibinfo {year} {2002})}\BibitemShut
  {NoStop}%
\bibitem [{\citenamefont {{Schober}}\ \emph
  {et~al.}(2012{\natexlab{a}})\citenamefont {{Schober}}, \citenamefont
  {{Schleicher}}, \citenamefont {{Federrath}}, \citenamefont {{Klessen}},\ and\
  \citenamefont {{Banerjee}}}]{SchoberEtAl2012.1}%
  \BibitemOpen
  \bibfield  {author} {\bibinfo {author} {\bibfnamefont {J.}~\bibnamefont
  {{Schober}}}, \bibinfo {author} {\bibfnamefont {D.}~\bibnamefont
  {{Schleicher}}}, \bibinfo {author} {\bibfnamefont {C.}~\bibnamefont
  {{Federrath}}}, \bibinfo {author} {\bibfnamefont {R.}~\bibnamefont
  {{Klessen}}}, \ and\ \bibinfo {author} {\bibfnamefont {R.}~\bibnamefont
  {{Banerjee}}},\ }\href {\doibase 10.1103/PhysRevE.85.026303} {\bibfield
  {journal} {\bibinfo  {journal} {\pre}\ }\textbf {\bibinfo {volume} {85}},\
  \bibinfo {eid} {026303} (\bibinfo {year} {2012}{\natexlab{a}})}\BibitemShut
  {NoStop}%
\bibitem [{\citenamefont {{Schekochihin}}\ \emph {et~al.}(2005)\citenamefont
  {{Schekochihin}}, \citenamefont {{Haugen}}, \citenamefont {{Brandenburg}},
  \citenamefont {{Cowley}}, \citenamefont {{Maron}},\ and\ \citenamefont
  {{McWilliams}}}]{SchekochihinEtAl2005}%
  \BibitemOpen
  \bibfield  {author} {\bibinfo {author} {\bibfnamefont {A.~A.}\ \bibnamefont
  {{Schekochihin}}}, \bibinfo {author} {\bibfnamefont {N.~E.~L.}\ \bibnamefont
  {{Haugen}}}, \bibinfo {author} {\bibfnamefont {A.}~\bibnamefont
  {{Brandenburg}}}, \bibinfo {author} {\bibfnamefont {S.~C.}\ \bibnamefont
  {{Cowley}}}, \bibinfo {author} {\bibfnamefont {J.~L.}\ \bibnamefont
  {{Maron}}}, \ and\ \bibinfo {author} {\bibfnamefont {J.~C.}\ \bibnamefont
  {{McWilliams}}},\ }\href {\doibase 10.1086/431214} {\bibfield  {journal}
  {\bibinfo  {journal} {\apjl}\ }\textbf {\bibinfo {volume} {625}},\ \bibinfo
  {pages} {L115} (\bibinfo {year} {2005})}\BibitemShut {NoStop}%
\bibitem [{\citenamefont {{Schekochihin}}\ \emph {et~al.}(2007)\citenamefont
  {{Schekochihin}}, \citenamefont {{Iskakov}}, \citenamefont {{Cowley}},
  \citenamefont {{McWilliams}}, \citenamefont {{Proctor}},\ and\ \citenamefont
  {{Yousef}}}]{SchekochihinEtAl2007}%
  \BibitemOpen
  \bibfield  {author} {\bibinfo {author} {\bibfnamefont {A.~A.}\ \bibnamefont
  {{Schekochihin}}}, \bibinfo {author} {\bibfnamefont {A.~B.}\ \bibnamefont
  {{Iskakov}}}, \bibinfo {author} {\bibfnamefont {S.~C.}\ \bibnamefont
  {{Cowley}}}, \bibinfo {author} {\bibfnamefont {J.~C.}\ \bibnamefont
  {{McWilliams}}}, \bibinfo {author} {\bibfnamefont {M.~R.~E.}\ \bibnamefont
  {{Proctor}}}, \ and\ \bibinfo {author} {\bibfnamefont {T.~A.}\ \bibnamefont
  {{Yousef}}},\ }\href {\doibase 10.1088/1367-2630/9/8/300} {\bibfield
  {journal} {\bibinfo  {journal} {New Journal of Physics}\ }\textbf {\bibinfo
  {volume} {9}},\ \bibinfo {pages} {300} (\bibinfo {year} {2007})}\BibitemShut
  {NoStop}%
\bibitem [{\citenamefont {Malyshkin}\ and\ \citenamefont
  {Boldyrev}(2010)}]{MalyshkinBoldyrev2010}%
  \BibitemOpen
  \bibfield  {author} {\bibinfo {author} {\bibfnamefont {L.~M.}\ \bibnamefont
  {Malyshkin}}\ and\ \bibinfo {author} {\bibfnamefont {S.}~\bibnamefont
  {Boldyrev}},\ }\href {\doibase 10.1103/PhysRevLett.105.215002} {\bibfield
  {journal} {\bibinfo  {journal} {Phys. Rev. Lett.}\ }\textbf {\bibinfo
  {volume} {105}},\ \bibinfo {pages} {215002} (\bibinfo {year}
  {2010})}\BibitemShut {NoStop}%
\bibitem [{\citenamefont {{Kleeorin}}\ and\ \citenamefont
  {{Rogachevskii}}(2012)}]{KleeorinRogachevskii2012}%
  \BibitemOpen
  \bibfield  {author} {\bibinfo {author} {\bibfnamefont {N.}~\bibnamefont
  {{Kleeorin}}}\ and\ \bibinfo {author} {\bibfnamefont {I.}~\bibnamefont
  {{Rogachevskii}}},\ }\href {\doibase 10.1088/0031-8949/86/01/018404}
  {\bibfield  {journal} {\bibinfo  {journal} {Physica Scripta}\ }\textbf
  {\bibinfo {volume} {86}},\ \bibinfo {pages} {018404} (\bibinfo {year}
  {2012})}\BibitemShut {NoStop}%
\bibitem [{\citenamefont {Federrath}\ \emph {et~al.}(2011)\citenamefont
  {Federrath}, \citenamefont {Chabrier}, \citenamefont {Schober}, \citenamefont
  {Banerjee}, \citenamefont {Klessen},\ and\ \citenamefont
  {Schleicher}}]{FederrathEtAl2011.2}%
  \BibitemOpen
  \bibfield  {author} {\bibinfo {author} {\bibfnamefont {C.}~\bibnamefont
  {Federrath}}, \bibinfo {author} {\bibfnamefont {G.}~\bibnamefont {Chabrier}},
  \bibinfo {author} {\bibfnamefont {J.}~\bibnamefont {Schober}}, \bibinfo
  {author} {\bibfnamefont {R.}~\bibnamefont {Banerjee}}, \bibinfo {author}
  {\bibfnamefont {R.~S.}\ \bibnamefont {Klessen}}, \ and\ \bibinfo {author}
  {\bibfnamefont {D.~R.~G.}\ \bibnamefont {Schleicher}},\ }\href {\doibase
  10.1103/PhysRevLett.107.114504} {\bibfield  {journal} {\bibinfo  {journal}
  {Phys. Rev. Lett.}\ }\textbf {\bibinfo {volume} {107}},\ \bibinfo {pages}
  {114504} (\bibinfo {year} {2011})}\BibitemShut {NoStop}%
\bibitem [{\citenamefont {{Schober}}\ \emph
  {et~al.}(2012{\natexlab{b}})\citenamefont {{Schober}}, \citenamefont
  {{Schleicher}}, \citenamefont {{Federrath}}, \citenamefont {{Glover}},
  \citenamefont {{Klessen}},\ and\ \citenamefont
  {{Banerjee}}}]{SchoberEtAl2012.2}%
  \BibitemOpen
  \bibfield  {author} {\bibinfo {author} {\bibfnamefont {J.}~\bibnamefont
  {{Schober}}}, \bibinfo {author} {\bibfnamefont {D.}~\bibnamefont
  {{Schleicher}}}, \bibinfo {author} {\bibfnamefont {C.}~\bibnamefont
  {{Federrath}}}, \bibinfo {author} {\bibfnamefont {S.}~\bibnamefont
  {{Glover}}}, \bibinfo {author} {\bibfnamefont {R.~S.}\ \bibnamefont
  {{Klessen}}}, \ and\ \bibinfo {author} {\bibfnamefont {R.}~\bibnamefont
  {{Banerjee}}},\ }\href {\doibase 10.1088/0004-637X/754/2/99} {\bibfield
  {journal} {\bibinfo  {journal} {\apj}\ }\textbf {\bibinfo {volume} {754}},\
  \bibinfo {eid} {99} (\bibinfo {year} {2012}{\natexlab{b}})}\BibitemShut
  {NoStop}%
\bibitem [{\citenamefont {Schekochihin}\ \emph {et~al.}(2004)\citenamefont
  {Schekochihin}, \citenamefont {Cowley}, \citenamefont {Maron},\ and\
  \citenamefont {McWilliams}}]{SchekochihinEtAl2004}%
  \BibitemOpen
  \bibfield  {author} {\bibinfo {author} {\bibfnamefont {A.~A.}\ \bibnamefont
  {Schekochihin}}, \bibinfo {author} {\bibfnamefont {S.~C.}\ \bibnamefont
  {Cowley}}, \bibinfo {author} {\bibfnamefont {J.~L.}\ \bibnamefont {Maron}}, \
  and\ \bibinfo {author} {\bibfnamefont {J.~C.}\ \bibnamefont {McWilliams}},\
  }\href {\doibase 10.1103/PhysRevLett.92.054502} {\bibfield  {journal}
  {\bibinfo  {journal} {Phys. Rev. Lett.}\ }\textbf {\bibinfo {volume} {92}},\
  \bibinfo {pages} {054502} (\bibinfo {year} {2004})}\BibitemShut {NoStop}%
\bibitem [{\citenamefont {Roberts}\ and\ \citenamefont
  {Glatzmaier}(2000)}]{RobertsGlatzmaier2000}%
  \BibitemOpen
  \bibfield  {author} {\bibinfo {author} {\bibfnamefont {P.~H.}\ \bibnamefont
  {Roberts}}\ and\ \bibinfo {author} {\bibfnamefont {G.~A.}\ \bibnamefont
  {Glatzmaier}},\ }\href {\doibase 10.1103/RevModPhys.72.1081} {\bibfield
  {journal} {\bibinfo  {journal} {Rev. Mod. Phys.}\ }\textbf {\bibinfo {volume}
  {72}},\ \bibinfo {pages} {1081} (\bibinfo {year} {2000})}\BibitemShut
  {NoStop}%
\bibitem [{\citenamefont {{Parker}}(1971)}]{Parker1971}%
  \BibitemOpen
  \bibfield  {author} {\bibinfo {author} {\bibfnamefont {E.~N.}\ \bibnamefont
  {{Parker}}},\ }\href {\doibase 10.1086/150862} {\bibfield  {journal}
  {\bibinfo  {journal} {\apj}\ }\textbf {\bibinfo {volume} {164}},\ \bibinfo
  {pages} {491} (\bibinfo {year} {1971})}\BibitemShut {NoStop}%
\bibitem [{\citenamefont {{Parker}}(1975)}]{Parker1975}%
  \BibitemOpen
  \bibfield  {author} {\bibinfo {author} {\bibfnamefont {E.~N.}\ \bibnamefont
  {{Parker}}},\ }\href {\doibase 10.1086/153593} {\bibfield  {journal}
  {\bibinfo  {journal} {\apj}\ }\textbf {\bibinfo {volume} {198}},\ \bibinfo
  {pages} {205} (\bibinfo {year} {1975})}\BibitemShut {NoStop}%
\bibitem [{\citenamefont {{Brandenburg}}\ and\ \citenamefont
  {{Subramanian}}(2005)}]{BrandenburgSubramanian2005}%
  \BibitemOpen
  \bibfield  {author} {\bibinfo {author} {\bibfnamefont {A.}~\bibnamefont
  {{Brandenburg}}}\ and\ \bibinfo {author} {\bibfnamefont {K.}~\bibnamefont
  {{Subramanian}}},\ }\href {\doibase 10.1016/j.physrep.2005.06.005} {\bibfield
   {journal} {\bibinfo  {journal} {\physrep}\ }\textbf {\bibinfo {volume}
  {417}},\ \bibinfo {pages} {1} (\bibinfo {year} {2005})}\BibitemShut {NoStop}%
\bibitem [{\citenamefont {{Miesch}}(2012)}]{Miesch2012}%
  \BibitemOpen
  \bibfield  {author} {\bibinfo {author} {\bibfnamefont {M.~S.}\ \bibnamefont
  {{Miesch}}},\ }\href {\doibase 10.1098/rsta.2011.0507} {\bibfield  {journal}
  {\bibinfo  {journal} {Royal Society of London Philosophical Transactions
  Series A}\ }\textbf {\bibinfo {volume} {370}},\ \bibinfo {pages} {3049}
  (\bibinfo {year} {2012})}\BibitemShut {NoStop}%
\bibitem [{\citenamefont {{Cattaneo}}, \citenamefont {{Emonet}},\ and\
  \citenamefont {{Weiss}}(2003)}]{CattaneoEtAl2003}%
  \BibitemOpen
  \bibfield  {author} {\bibinfo {author} {\bibfnamefont {F.}~\bibnamefont
  {{Cattaneo}}}, \bibinfo {author} {\bibfnamefont {T.}~\bibnamefont
  {{Emonet}}}, \ and\ \bibinfo {author} {\bibfnamefont {N.}~\bibnamefont
  {{Weiss}}},\ }\href {\doibase 10.1086/374313} {\bibfield  {journal} {\bibinfo
   {journal} {\apj}\ }\textbf {\bibinfo {volume} {588}},\ \bibinfo {pages}
  {1183} (\bibinfo {year} {2003})}\BibitemShut {NoStop}%
\bibitem [{\citenamefont {{Pipin}}\ and\ \citenamefont
  {{Seehafer}}(2009)}]{PipinSeehafer2009}%
  \BibitemOpen
  \bibfield  {author} {\bibinfo {author} {\bibfnamefont {V.~V.}\ \bibnamefont
  {{Pipin}}}\ and\ \bibinfo {author} {\bibfnamefont {N.}~\bibnamefont
  {{Seehafer}}},\ }\href {\doibase 10.1051/0004-6361:200810766} {\bibfield
  {journal} {\bibinfo  {journal} {\aap}\ }\textbf {\bibinfo {volume} {493}},\
  \bibinfo {pages} {819} (\bibinfo {year} {2009})}\BibitemShut {NoStop}%
\bibitem [{\citenamefont {Nornberg}\ \emph {et~al.}(2006)\citenamefont
  {Nornberg}, \citenamefont {Spence}, \citenamefont {Kendrick}, \citenamefont
  {Jacobson},\ and\ \citenamefont {Forest}}]{NornbergEtAl2006}%
  \BibitemOpen
  \bibfield  {author} {\bibinfo {author} {\bibfnamefont {M.~D.}\ \bibnamefont
  {Nornberg}}, \bibinfo {author} {\bibfnamefont {E.~J.}\ \bibnamefont
  {Spence}}, \bibinfo {author} {\bibfnamefont {R.~D.}\ \bibnamefont
  {Kendrick}}, \bibinfo {author} {\bibfnamefont {C.~M.}\ \bibnamefont
  {Jacobson}}, \ and\ \bibinfo {author} {\bibfnamefont {C.~B.}\ \bibnamefont
  {Forest}},\ }\href {\doibase 10.1103/PhysRevLett.97.044503} {\bibfield
  {journal} {\bibinfo  {journal} {Phys. Rev. Lett.}\ }\textbf {\bibinfo
  {volume} {97}},\ \bibinfo {pages} {044503} (\bibinfo {year}
  {2006})}\BibitemShut {NoStop}%
\bibitem [{\citenamefont {{Kolmogorov}}(1941)}]{Kolmogorov1941}%
  \BibitemOpen
  \bibfield  {author} {\bibinfo {author} {\bibfnamefont {A.}~\bibnamefont
  {{Kolmogorov}}},\ }\href@noop {} {\bibfield  {journal} {\bibinfo  {journal}
  {Akademiia Nauk SSSR Doklady}\ }\textbf {\bibinfo {volume} {30}},\ \bibinfo
  {pages} {301} (\bibinfo {year} {1941})}\BibitemShut {NoStop}%
\bibitem [{\citenamefont {{Mac Low}}\ and\ \citenamefont
  {{Klessen}}(2004)}]{MacLowKlessen2004}%
  \BibitemOpen
  \bibfield  {author} {\bibinfo {author} {\bibfnamefont {M.-M.}\ \bibnamefont
  {{Mac Low}}}\ and\ \bibinfo {author} {\bibfnamefont {R.~S.}\ \bibnamefont
  {{Klessen}}},\ }\href {\doibase 10.1103/RevModPhys.76.125} {\bibfield
  {journal} {\bibinfo  {journal} {Rev. Mod. Phys.}\ }\textbf {\bibinfo {volume}
  {76}},\ \bibinfo {pages} {125} (\bibinfo {year} {2004})}\BibitemShut
  {NoStop}%
\bibitem [{\citenamefont {{Elmegreen}}\ and\ \citenamefont
  {{Scalo}}(2004)}]{ElmegreenScalo2004}%
  \BibitemOpen
  \bibfield  {author} {\bibinfo {author} {\bibfnamefont {B.~G.}\ \bibnamefont
  {{Elmegreen}}}\ and\ \bibinfo {author} {\bibfnamefont {J.}~\bibnamefont
  {{Scalo}}},\ }\href {\doibase 10.1146/annurev.astro.41.011802.094859}
  {\bibfield  {journal} {\bibinfo  {journal} {\araa}\ }\textbf {\bibinfo
  {volume} {42}},\ \bibinfo {pages} {211} (\bibinfo {year} {2004})}\BibitemShut
  {NoStop}%
\bibitem [{\citenamefont {{Scalo}}\ and\ \citenamefont
  {{Elmegreen}}(2004)}]{ScaloElmegreen2004}%
  \BibitemOpen
  \bibfield  {author} {\bibinfo {author} {\bibfnamefont {J.}~\bibnamefont
  {{Scalo}}}\ and\ \bibinfo {author} {\bibfnamefont {B.~G.}\ \bibnamefont
  {{Elmegreen}}},\ }\href@noop {} {\bibfield  {journal} {\bibinfo  {journal}
  {\araa}\ }\textbf {\bibinfo {volume} {42}} (\bibinfo {year}
  {2004})}\BibitemShut {NoStop}%
\bibitem [{\citenamefont {{Hennebelle}}\ \emph {et~al.}(2008)\citenamefont
  {{Hennebelle}}, \citenamefont {{Banerjee}}, \citenamefont
  {{V{\'a}zquez-Semadeni}}, \citenamefont {{Klessen}},\ and\ \citenamefont
  {{Audit}}}]{HennebelleEtAl2008}%
  \BibitemOpen
  \bibfield  {author} {\bibinfo {author} {\bibfnamefont {P.}~\bibnamefont
  {{Hennebelle}}}, \bibinfo {author} {\bibfnamefont {R.}~\bibnamefont
  {{Banerjee}}}, \bibinfo {author} {\bibfnamefont {E.}~\bibnamefont
  {{V{\'a}zquez-Semadeni}}}, \bibinfo {author} {\bibfnamefont {R.~S.}\
  \bibnamefont {{Klessen}}}, \ and\ \bibinfo {author} {\bibfnamefont
  {E.}~\bibnamefont {{Audit}}},\ }\href {\doibase 10.1051/0004-6361:200810165}
  {\bibfield  {journal} {\bibinfo  {journal} {\aap}\ }\textbf {\bibinfo
  {volume} {486}},\ \bibinfo {pages} {L43} (\bibinfo {year}
  {2008})}\BibitemShut {NoStop}%
\bibitem [{\citenamefont {{Banerjee}}\ \emph {et~al.}(2009)\citenamefont
  {{Banerjee}}, \citenamefont {{V{\'a}zquez-Semadeni}}, \citenamefont
  {{Hennebelle}},\ and\ \citenamefont {{Klessen}}}]{BanerjeeEtAl2009}%
  \BibitemOpen
  \bibfield  {author} {\bibinfo {author} {\bibfnamefont {R.}~\bibnamefont
  {{Banerjee}}}, \bibinfo {author} {\bibfnamefont {E.}~\bibnamefont
  {{V{\'a}zquez-Semadeni}}}, \bibinfo {author} {\bibfnamefont {P.}~\bibnamefont
  {{Hennebelle}}}, \ and\ \bibinfo {author} {\bibfnamefont {R.~S.}\
  \bibnamefont {{Klessen}}},\ }\href {\doibase
  10.1111/j.1365-2966.2009.15115.x} {\bibfield  {journal} {\bibinfo  {journal}
  {\mnras}\ }\textbf {\bibinfo {volume} {398}},\ \bibinfo {pages} {1082}
  (\bibinfo {year} {2009})}\BibitemShut {NoStop}%
\bibitem [{\citenamefont {{Klessen}}\ and\ \citenamefont
  {{Hennebelle}}(2010)}]{KlessenHennebelle2010}%
  \BibitemOpen
  \bibfield  {author} {\bibinfo {author} {\bibfnamefont {R.~S.}\ \bibnamefont
  {{Klessen}}}\ and\ \bibinfo {author} {\bibfnamefont {P.}~\bibnamefont
  {{Hennebelle}}},\ }\href {\doibase 10.1051/0004-6361/200913780} {\bibfield
  {journal} {\bibinfo  {journal} {\aap}\ }\textbf {\bibinfo {volume} {520}},\
  \bibinfo {pages} {A17+} (\bibinfo {year} {2010})}\BibitemShut {NoStop}%
\bibitem [{\citenamefont {{Larson}}(1981)}]{Larson1981}%
  \BibitemOpen
  \bibfield  {author} {\bibinfo {author} {\bibfnamefont {R.~B.}\ \bibnamefont
  {{Larson}}},\ }\href@noop {} {\bibfield  {journal} {\bibinfo  {journal}
  {\mnras}\ }\textbf {\bibinfo {volume} {194}},\ \bibinfo {pages} {809}
  (\bibinfo {year} {1981})}\BibitemShut {NoStop}%
\bibitem [{\citenamefont {{She}}\ and\ \citenamefont
  {{Leveque}}(1994)}]{SheLeveque1994}%
  \BibitemOpen
  \bibfield  {author} {\bibinfo {author} {\bibfnamefont {Z.-S.}\ \bibnamefont
  {{She}}}\ and\ \bibinfo {author} {\bibfnamefont {E.}~\bibnamefont
  {{Leveque}}},\ }\href {\doibase 10.1103/PhysRevLett.72.336} {\bibfield
  {journal} {\bibinfo  {journal} {Phys. Rev. Lett.}\ }\textbf {\bibinfo
  {volume} {72}},\ \bibinfo {pages} {336} (\bibinfo {year} {1994})}\BibitemShut
  {NoStop}%
\bibitem [{\citenamefont {{Boldyrev}}, \citenamefont {{Nordlund}},\ and\
  \citenamefont {{Padoan}}(2002)}]{Boldyrev2002}%
  \BibitemOpen
  \bibfield  {author} {\bibinfo {author} {\bibfnamefont {S.}~\bibnamefont
  {{Boldyrev}}}, \bibinfo {author} {\bibfnamefont {{\AA}.}~\bibnamefont
  {{Nordlund}}}, \ and\ \bibinfo {author} {\bibfnamefont {P.}~\bibnamefont
  {{Padoan}}},\ }\href {\doibase 10.1086/340758} {\bibfield  {journal}
  {\bibinfo  {journal} {\apj}\ }\textbf {\bibinfo {volume} {573}},\ \bibinfo
  {pages} {678} (\bibinfo {year} {2002})}\BibitemShut {NoStop}%
\bibitem [{\citenamefont {{Federrath}}\ \emph {et~al.}(2010)\citenamefont
  {{Federrath}}, \citenamefont {{Roman-Duval}}, \citenamefont {{Klessen}},
  \citenamefont {{Schmidt}},\ and\ \citenamefont {{Mac Low}}}]{Federrath2010}%
  \BibitemOpen
  \bibfield  {author} {\bibinfo {author} {\bibfnamefont {C.}~\bibnamefont
  {{Federrath}}}, \bibinfo {author} {\bibfnamefont {J.}~\bibnamefont
  {{Roman-Duval}}}, \bibinfo {author} {\bibfnamefont {R.~S.}\ \bibnamefont
  {{Klessen}}}, \bibinfo {author} {\bibfnamefont {W.}~\bibnamefont
  {{Schmidt}}}, \ and\ \bibinfo {author} {\bibfnamefont {M.-M.}\ \bibnamefont
  {{Mac Low}}},\ }\href {\doibase 10.1051/0004-6361/200912437} {\bibfield
  {journal} {\bibinfo  {journal} {\aap}\ }\textbf {\bibinfo {volume} {512}},\
  \bibinfo {pages} {A81} (\bibinfo {year} {2010})}\BibitemShut {NoStop}%
\bibitem [{\citenamefont {{Ossenkopf}}\ and\ \citenamefont {{Mac
  Low}}(2002)}]{OssenkopfMacLow2002}%
  \BibitemOpen
  \bibfield  {author} {\bibinfo {author} {\bibfnamefont {V.}~\bibnamefont
  {{Ossenkopf}}}\ and\ \bibinfo {author} {\bibfnamefont {M.-M.}\ \bibnamefont
  {{Mac Low}}},\ }\href {\doibase 10.1051/0004-6361:20020629} {\bibfield
  {journal} {\bibinfo  {journal} {\aap}\ }\textbf {\bibinfo {volume} {390}},\
  \bibinfo {pages} {307} (\bibinfo {year} {2002})}\BibitemShut {NoStop}%
\bibitem [{\citenamefont {Burgers}(1948)}]{Burgers1948}%
  \BibitemOpen
  \bibfield  {author} {\bibinfo {author} {\bibfnamefont {J.}~\bibnamefont
  {Burgers}},\ }\href {\doibase DOI: 10.1016/S0065-2156(08)70100-5} {\emph
  {\bibinfo {title} {{A Mathematical Model Illustrating the Theory of
  Turbulence}}}},\ \bibinfo {series} {Advances in Applied Mechanics},
  Vol.~\bibinfo {volume} {1}\ (\bibinfo  {publisher} {Elsevier},\ \bibinfo
  {year} {1948})\ pp.\ \bibinfo {pages} {171 -- 199}\BibitemShut {NoStop}%
\bibitem [{\citenamefont {{Vainshtein}}\ and\ \citenamefont
  {{Zeldovich}}(1972)}]{VainshteinZeldovich1974}%
  \BibitemOpen
  \bibfield  {author} {\bibinfo {author} {\bibfnamefont {S.~I.}\ \bibnamefont
  {{Vainshtein}}}\ and\ \bibinfo {author} {\bibfnamefont {Y.~B.}\ \bibnamefont
  {{Zeldovich}}},\ }\href@noop {} {\bibfield  {journal} {\bibinfo  {journal}
  {Sov.~Phys.~Usp.}\ }\textbf {\bibinfo {volume} {15}} (\bibinfo {year}
  {1972})}\BibitemShut {NoStop}%
\bibitem [{\citenamefont {Batchelor}(1953)}]{Bachelor1953}%
  \BibitemOpen
  \bibfield  {author} {\bibinfo {author} {\bibfnamefont {G.~K.}\ \bibnamefont
  {Batchelor}},\ }\href@noop {} {\emph {\bibinfo {title} {{The theory of
  homogeneous turbulence}}}},\ Cambridge monographs on mechanics and applied
  mathematics\ (\bibinfo  {publisher} {Cambridge University Press},\ \bibinfo
  {address} {Cambridge},\ \bibinfo {year} {1953})\ pp.\ \bibinfo {pages} {XI,
  197 S.}\BibitemShut {Stop}%
\bibitem [{\citenamefont {{Schmidt}}\ \emph {et~al.}(2009)\citenamefont
  {{Schmidt}}, \citenamefont {{Federrath}}, \citenamefont {{Hupp}},
  \citenamefont {{Kern}},\ and\ \citenamefont {{Niemeyer}}}]{Schmidt2009}%
  \BibitemOpen
  \bibfield  {author} {\bibinfo {author} {\bibfnamefont {W.}~\bibnamefont
  {{Schmidt}}}, \bibinfo {author} {\bibfnamefont {C.}~\bibnamefont
  {{Federrath}}}, \bibinfo {author} {\bibfnamefont {M.}~\bibnamefont {{Hupp}}},
  \bibinfo {author} {\bibfnamefont {S.}~\bibnamefont {{Kern}}}, \ and\ \bibinfo
  {author} {\bibfnamefont {J.~C.}\ \bibnamefont {{Niemeyer}}},\ }\href
  {\doibase 10.1051/0004-6361:200809967} {\bibfield  {journal} {\bibinfo
  {journal} {\aap}\ }\textbf {\bibinfo {volume} {494}},\ \bibinfo {pages} {127}
  (\bibinfo {year} {2009})}\BibitemShut {NoStop}%
\bibitem [{Note1()}]{Note1}%
  \BibitemOpen
  \bibinfo {note} {We note that there is a typo in the paper of \protect \citet
  {SchoberEtAl2012.1}, where the potential was derived for a general type of
  turbulence. The term $2\kappa _\protect \text {diff}$ / $r^2$ appeared here
  twice.}\BibitemShut {Stop}%
\bibitem [{\citenamefont {{Mestel}}\ and\ \citenamefont
  {{Subramanian}}(1991)}]{MestelSubramanian1991}%
  \BibitemOpen
  \bibfield  {author} {\bibinfo {author} {\bibfnamefont {L.}~\bibnamefont
  {{Mestel}}}\ and\ \bibinfo {author} {\bibfnamefont {K.}~\bibnamefont
  {{Subramanian}}},\ }\href@noop {} {\bibfield  {journal} {\bibinfo  {journal}
  {\mnras}\ }\textbf {\bibinfo {volume} {248}},\ \bibinfo {pages} {677}
  (\bibinfo {year} {1991})}\BibitemShut {NoStop}%
\bibitem [{\citenamefont {Boldyrev}\ and\ \citenamefont
  {Cattaneo}(2004)}]{BoldyrevCattaneo2004}%
  \BibitemOpen
  \bibfield  {author} {\bibinfo {author} {\bibfnamefont {S.}~\bibnamefont
  {Boldyrev}}\ and\ \bibinfo {author} {\bibfnamefont {F.}~\bibnamefont
  {Cattaneo}},\ }\href {\doibase 10.1103/PhysRevLett.92.144501} {\bibfield
  {journal} {\bibinfo  {journal} {Phys. Rev. Lett.}\ }\textbf {\bibinfo
  {volume} {92}},\ \bibinfo {pages} {144501} (\bibinfo {year}
  {2004})}\BibitemShut {NoStop}%
\bibitem [{\citenamefont {{Iskakov}}\ \emph {et~al.}(2007)\citenamefont
  {{Iskakov}}, \citenamefont {{Schekochihin}}, \citenamefont {{Cowley}},
  \citenamefont {{McWilliams}},\ and\ \citenamefont
  {{Proctor}}}]{IskakovEtAl2007}%
  \BibitemOpen
  \bibfield  {author} {\bibinfo {author} {\bibfnamefont {A.~B.}\ \bibnamefont
  {{Iskakov}}}, \bibinfo {author} {\bibfnamefont {A.~A.}\ \bibnamefont
  {{Schekochihin}}}, \bibinfo {author} {\bibfnamefont {S.~C.}\ \bibnamefont
  {{Cowley}}}, \bibinfo {author} {\bibfnamefont {J.~C.}\ \bibnamefont
  {{McWilliams}}}, \ and\ \bibinfo {author} {\bibfnamefont {M.~R.~E.}\
  \bibnamefont {{Proctor}}},\ }\href {\doibase 10.1103/PhysRevLett.98.208501}
  {\bibfield  {journal} {\bibinfo  {journal} {Physical Review Letters}\
  }\textbf {\bibinfo {volume} {98}},\ \bibinfo {eid} {208501} (\bibinfo {year}
  {2007})}\BibitemShut {NoStop}%
\bibitem [{\citenamefont {{Bovino}}, \citenamefont {{Schleicher}},\ and\
  \citenamefont {{Schober}}(2012)}]{BovinoSchleicherSchober2012}%
  \BibitemOpen
  \bibfield  {author} {\bibinfo {author} {\bibfnamefont {S.}~\bibnamefont
  {{Bovino}}}, \bibinfo {author} {\bibfnamefont {D.~R.~G.}\ \bibnamefont
  {{Schleicher}}}, \ and\ \bibinfo {author} {\bibfnamefont {J.}~\bibnamefont
  {{Schober}}},\ }\href@noop {} {\bibfield  {journal} {\bibinfo  {journal}
  {ArXiv e-prints}\ } (\bibinfo {year} {2012})},\ \bibinfo {note}
  {arXiv:astro-ph/1212.3419}\BibitemShut {NoStop}%
\end{thebibliography}

%

\end{document}